\documentclass[final,11pt]{elsarticle}
\usepackage[latin1]{inputenc}
\usepackage{amsmath,bm}
\usepackage{amsthm}
\usepackage{amsfonts}
\usepackage{color}
\usepackage[dvipsnames]{xcolor}
\usepackage{amssymb}
\usepackage{graphicx}
\usepackage{comment}
\usepackage{caption}
\usepackage{subcaption}
\usepackage{comment,booktabs,multirow}
\usepackage[ruled]{algorithm}
\usepackage{empheq}
\usepackage[noend]{algpseudocode}
\usepackage[top=1in, bottom=1in, left=1in, right=1in]{geometry}

\usepackage[pdftex, unicode=true,
            colorlinks,
            linkcolor=red,
            anchorcolor=blue,
            citecolor=green
            ]{hyperref}

\newtheorem{thm}{Theorem}[section]

\newtheorem{remark}[thm]{Remark}
\newtheorem{proposition}{Proposition}

\newcommand{\lb}{\left (}
\newcommand{\rb}{\right )}

\newcommand{\be}{\bm e}

\newcommand{\bn}{\bm n}

\newcommand{\bW}{\bm W}
\newcommand{\bx}{\bm x}

\newcommand{\bX}{\bm X}

\newcommand{\bu}{\bm u}

\newcommand{\bv}{\bm v}

\newcommand{\bV}{\bm V}

\newcommand{\bk}{\bm k}

\newcommand{\bD}{\bm D}

\newcommand{\bP}{\bm P}

\newcommand{\bT}{\bm T}

\DeclareMathOperator*{\argmin}{\arg\!\min}

\definecolor{orangeR}{RGB}{255,69,0}
\definecolor{purple}{rgb}{0.4,0.1,0.2}
\definecolor{lightgreen}{RGB}{152,251,152}
\definecolor{darkgreen}{RGB}{0,100,0}
\definecolor{lightorange}{rgb}{1, 0.63,0.48}
\definecolor{lightblue}{rgb}{0.53, 0.81,0.99}
\definecolor{navy}{rgb}{0,0,0.5}
\definecolor{darkcyan}{RGB}{0,139,139}
\definecolor{palatinateblue}{rgb}{0.15, 0.23, 0.89}
\definecolor{freeblue}{rgb}{0.25,0.41,0.88}
\definecolor{brown}{RGB}{165,42,42}

\makeatletter
\newcounter{manualsubequation}
\renewcommand{\themanualsubequation}{\alph{manualsubequation}}
\newcommand{\startsubequation}{%
	\setcounter{manualsubequation}{0}%
	\refstepcounter{equation}\ltx@label{manualsubeq\theequation}%
	\xdef\labelfor@subeq{manualsubeq\theequation}%
}
\newcommand{\tagsubequation}{%
	\stepcounter{manualsubequation}%
	\tag{\ref{\labelfor@subeq}\themanualsubequation}%
}
\let\subequationlabel\ltx@label
\makeatother

\makeatletter
\def\ps@pprintTitle{%
 \let\@oddhead\@empty
 \let\@evenhead\@empty
 \def\@oddfoot{}%
 \let\@evenfoot\@oddfoot}
\makeatother

\begin{document}
    \begin{frontmatter}
    \title{Parallel Exponential Time Differencing Methods for Geophysical Flow Simulations}
\author[usc]{Rihui Lan}
\ead{rlan@mailbox.sc.edu}
\author[cas]{Wei Leng}
\ead{wleng@lsec.cc.ac.cn}
\author[usc]{Zhu Wang}
\ead{wangzhu@math.sc.edu}
\author[usc]{Lili Ju\corref{cor}}
\ead{ju@math.sc.edu}
\author[fsu]{Max Gunzburger}
\ead{mgunzburger@fsu.edu}
\address[usc]{Department of Mathematics, University of South Carolina, Columbia, SC 29208}
\address[cas]{State Key Laboratory of Scientific and Engineering Computing, Chinese Academy of Sciences,
	Beijing 100190, China}
\address[fsu]{Department of Scientific Computing, Florida State University, Tallahassee, FL 32306}
\cortext[cor]{Corresponding author}

\begin{abstract}
Two ocean models are considered for geophysical flow simulations: the multi-layer shallow water equations and the multi-layer primitive equations. For the former, we investigate the parallel performance of exponential time differencing (ETD) methods, including  exponential Rosenbrock-Euler, ETD2wave, and B-ETD2wave. For the latter, we take advantage of the splitting of barotropic and baroclinic modes and propose a new two-level method in which an ETD method is applied to solve the fast barotropic mode.  These methods could  improve the computational efficiency of  numerical simulations because ETD methods allow for much larger time step sizes than traditional explicit time-stepping techniques that are commonly used in existing computational ocean  models. Several standard benchmark tests for ocean modeling are performed and comparison of the numerical results  demonstrate a great potential of applying the parallel ETD methods for simulating real-world geophysical flows.

\end{abstract}

\end{frontmatter}

\section{Introduction}
Without a doubt, the ocean is vitally important to everyone and significantly affects our daily life. To understand and predict ocean circulation, many numerical models have been developed based on fundamental laws of physics with a focus on the properties of geophysical flows. So far, the primitive equations, built upon the Boussinesq approximation, are widely used to model global or regional oceanic circulation, including the velocity, depth (or layer thickness), pressure, and tracers such as temperature, salinity, or chemicals. Due to the large aspect ratio, the ocean model can usually be simplified by regarding the ocean as a single-layer or a stack of immiscible layers for which each one is characterized by a constant density. The layered model is more accurate in representing vertical profiles and is often used to model the actual ocean. The reader is referred to \cite{cushman2011introduction} for additional details. 
On the other hand,  numerical simulations of the layered models become more challenging as the oceanic dynamical systems are of very large scales. In addition, oceanic flows are impacted by many other factors, including the coupling of external and internal gravity waves and planetary rotation which results in ocean dynamics involving multiple time scales. Therefore, to achieve accurate and stable numerical schemes, one has to choose small time step schemes such as in explicit Runge-Kutta methods, which have good parallel scalability and are commonly used in existing numerical ocean models.

To better handle the multiple time scales, some techniques based on the splitting strategies have been developed. For instance, in \cite{sandu2015generalized, gunther2016multirate}, G{\"u}nther and Sandu designed a generalized additive Runge-Kutta (GARK) method which allows fast and slow modes to use different stages and time step sizes. However, GARK needs extra effort devoted to the coupling conditions between different modes. In  \cite{bleck1990wind},
Bleck and Smith applied a splitting idea to predict the Atlantic ocean dynamics under the isopycnic coordinate system.
This approach separates the ocean motion into two modes: the barotropic (fast) mode and baroclinic (slow) mode. The fast mode is obtained via vertical averaging and the slow mode consists of the difference between the original velocity and the fast mode.  Based on Bleck and Smith's work, in \cite{higdon1997barotropic,higdon1999implementation,higdon2002two,higdon2005two}, Higdon et al. developed a two-level time-stepping method, one of the 
split-explicit time-stepping methods, to speed up the simulation . 
In their approach, the baroclinic mode is first advanced with a large time step size and after that, an iterative scheme is applied to determine the barotropic mode. However, the sub-stepping could still be time consuming. Therefore, to further improve the computational efficiency, in this work we  introduce a new two-level method that replaces the sub-stepping for the barotropic mode with an exponential time differencing (ETD) method. 

Exponential-integrator based methods, as an alternative to explicit or implicit time stepping, were introduced as early as the 1960s \cite{pope1963exponential,certaine1960solution} to solve  autonomous systems of first-order ordinary differential equations (ODEs). The systems are solved exactly via the variation of constants formulas or integrating factors, after which the temporal integrals of the matrix exponentials are approximated. These methods did not gain much attention until the early 1990's because of the difficulty encountered in evaluating exponential functions of matrices. Thanks to the progress in computer science and numerical linear algebra \cite{gs92,saad92,higham2005scaling, sidje1998expokit,moler2003nineteen,hochbruck1997krylov}, this method has received a renewed interest in solving various large systems of stiff semilinear or nonlinear systems \cite{cox2002exponential,hochbruck1998exponential,tokman2006efficient}. As for  its application to climate sciences, Clancy and Pudykiewicz investigated the potential of ETD methods in atmospheric models. They concluded that these schemes are more efficient than the semi-implicit methods by allowing far longer time-steps, but one needs to efficiently calculate the products of matrix exponentials and vectors. In \cite{gaudreault2016efficient}, Gaudreault and Pudykiewicz utilized Incomplete Orthogonalization
Method (IOM) to evaluate the matrix exponential functions. 
Recently, Pieper et al. applied the ETD method to speed up the simulations of the rotating multi-layer shallow water model in \cite{pieper2019exponential}. They reformulated the original equations into a Hamiltonian framework, then developed the ETD-wave and the barotropic ETD methods. By comparing the average simulated years per real day, they numerically showed the ETD methods are more efficient than the classical explicit Runge-Kutta methods. Whereas only sequential algorithms are addressed in those works, as we know, parallel computing is a powerful tool to accelerate scientific computing and have, necessarily, been widely used in numerical climate models. Hence, in this paper, we will further implement and investigate the performance of parallel ETD methods based on domain decomposition. In addition, there has not been much work applying  ETD methods to the primitive equations, with the most related research being that of Calandrini et al.\cite{calandrini2020exponential}, in which the ETD method is utilized to speed up the tracer equations in the primitive equations system. As mentioned earlier, we will  also attempt to replace the sub-steppings for the barotropic mode in the two-level time-stepping method with ETD.

In this paper we mainly focus on the parallel implementation of the ETD methods  for simulating two ocean models: the multi-layer shallow water equations and the multi-layer primitive equations. The computational settings and testbeds we used are taken from {\em MPAS-Ocean} \cite{petersen2013mpas}, a numerical ocean model developed at the Los Alamos National Laboratory for the simulation of the ocean system across scales in time and space.  
The rest of paper is outlined as follows. We first present the two types of  ocean dynamics equations in Section \ref{sect:dynamics}. For spatial discretization, we apply the TRiSK scheme that is briefly introduced in Section \ref{sect:trisk}. Section \ref{sect:etd} covers the illustration of different ETD methods for solving these two models and their parallel implementation. Numerical experiments are  performed and discussed  in  Section \ref{sect:experiments}. Finally, some conclusions are drawn in Section \ref{sect:conclusion}.

\section{Ocean dynamics and related models}\label{sect:dynamics}
In this section  we briefly review two widely-used ocean models: one is governed by the  multi-layer rotating shallow water equations (SWEs), the other is by  the multi-layer primitive equations. {For the case of shallow water equations, we  consider the equations reduced  by depth integration from the primitive equations and their variation in the framework of Hamiltonian systems developed in \cite{pieper2019exponential}. For the primitive equations, we specifically consider the ones used in {\em MPAS-Ocean} (\cite{petersen2013mpas}) which are the incompressible Boussinesq equations in hydrostatic balance and include several tracer equations.}

\subsection{Shallow water equations}
For geophysical flow with a vertical dimension much smaller than the horizontal scale, the shallow water equations  can be used to describe the fluid motion. This is a typical situation when fluid flows in the ocean, coastal regions, estuaries and rivers are concerned. 
The single-layer rotating shallow water model  is governed by: 
\begin{equation}\label{RSW:eq:single}
\left\{
\begin{split}
\partial_th&=-\nabla\cdot\lb h\bu\rb &\text{ in } \Omega,\\
\partial_t\bu&=-\nabla\lb K[\bu]+g(h+b)\rb-q[h,\bu]\hat{\bk}\times\lb h\bu\rb+\mathcal{G}\lb h,u\rb &\text{ in } \Omega,
\end{split}
\right.
\end{equation}
where $h= h(t,\bx)$ 
is the fluid thickness, $\bu = \bu(t,\bx)$ 
is the velocity tangential to the surface $\Omega$ (i.e., $\bu\cdot \hat{\bk}=0$) and the velocity is subject to zero normal flow boundary condition (i.e., $\bu\cdot\hat{n}=0$ on $\partial \Omega$). In addition, $K[\bu]=|\bu|^2/2$ is the kinetic energy, $\hat{\bk}\times\bu$ is the perpendicular velocity which is usually denoted by $\bu^{\perp}$, and $q[h,\bu]=(\hat{\bk}\cdot\nabla\times\bu+f)/h$ is the potential vorticity with $f$ being the Coriolis parameter. As discussed in \cite{pieper2019exponential}, the SWEs can be reformulated in the Hamiltonian framework by introducing the following Hamiltonian 
\begin{equation}
\mathcal{H}[h,\bu]=\int_{\Omega}\lb hK[\bu]+\frac{g}{2}(h+b)^2\rb \,{\rm d}{\bf x}.
\end{equation}
Let $V=\lb h,\bu\rb$ belong to the solution space $\mathcal{X}=L^2(\Omega)\times ({L^2(\Omega))^2}$, that is equipped with the inner product, for any $\phi = (\phi_h, \phi_{\bu}), \psi= \lb \psi_h, \psi_{\bu}\rb\in \mathcal{X}$, 
\begin{equation*}
\lb \phi, \psi\rb_{\mathcal{X}}=\int_{\Omega}\lb \phi_h \psi_h+\phi_{\bu}\cdot\psi_{\bu}\rb \,{\rm d}{\bf x}.
\end{equation*}
Then the functional derivative of $\mathcal{H}$   is given by 
\begin{equation}
\delta \mathcal{H}[V]=\frac{\delta \mathcal{H}}{\delta V}[V]=\begin{bmatrix}
K[\bu]+g(h+b)\\
h\bu
\end{bmatrix},
\end{equation}
and its directional derivative  is
\begin{equation}
\mathcal{H}'[V;W]=\lb \frac{\delta \mathcal{H}}{\delta V}[V],W\rb_{\mathcal{X}}=\int_{\Omega}\lb (K[\bu]+g(h+b))h_w+h\bu\cdot \bu_w\rb\,{\rm d}{\bf x}
\end{equation}
for any  $W=(h_w, \bu_w)$.
Define the skew-symmetric operator $\mathcal{J}$ as
\begin{equation}
\mathcal{J}[h,\bu]\coloneqq\begin{bmatrix}
0&-\nabla\cdot\\
-\nabla&-q[h,\bu]\hat{\bk}\times
\end{bmatrix}.
\end{equation}
Then the SWEs \eqref{RSW:eq:single} can be recast in the Hamiltonian framework as:
\begin{equation}\label{Hamilton:single}
\partial_t V=\mathcal{J}\delta\mathcal{H}[V]+\begin{bmatrix}
0\\
\mathcal{G}[h,\bu]
\end{bmatrix}. 
\end{equation}
\subsubsection{Multi-layer shallow water equations}
Besides the ambient rotation, stratification effects play an essential role in geophysical fluid dynamics. 
Layered models include the density differences that depict the stratified fluid as a finite number of moving layers, stacked one upon another, with an in-layer constant density. 
%
Such models can be derived by partitioning the vertical range into $L$ segments and imposing distinct density values in different layers that increase downward (i.e., $\rho_i<\rho_{i+1}, i=1,\dots,L-1$). The multi-layer SWEs in correspondence to the classic SWEs \eqref{RSW:eq:single}  reads: for $k=1,\dots,L$,
\begin{equation}\label{RSW:eq}
\left\{
\begin{split}
\partial_th_k&=-\nabla\cdot\lb h_k\bu_k\rb &\text{ in } \Omega,\\
\partial_t\bu_k&=-\nabla\lb K[\bu_k]+\lb g/\rho_k\rb p_k[h]\rb-q[h_k,\bu_k]\hat{\bk}\times\lb h_k\bu_k\rb+\mathcal{G}_k\lb h,\bu\rb &\text{ in } \Omega,
\end{split}
\right.
\end{equation}
where $h_k$ and $\bu_k$ are the thickness and velocity of layer $k$, and $K[\bu_k]$ and $q[h_k,\bu_k]$ are defined similarly as those in the single-layer case. 

Define $V= \lb h, \bu\rb\in \mathcal{X}^L$, the Cartesian production of $\mathcal{X}$ for $L$ times, which consists of  the variables of all layers. 
For any $\phi = (\phi_h, \phi_{\bu}), \psi= \lb \psi_h, \psi_{\bu}\rb\in \mathcal{X}^L$, the corresponding inner produce is defined by
\begin{equation*}
\lb \phi, \psi \rb_{\mathcal{X}^L}=\sum_{k=1}^{L}\int_{\Omega}\lb \phi_{k}^{h} \psi_{k}^{h}+\phi_{k}^{\bu}\cdot\psi_{k}^{\bu}\rb\,{\rm d}{\bf x}.
\end{equation*}
Different from the single-layer case, the pressure term $p_k[h]$ combines all the layers 
above layer $k$, defined as
\begin{equation}\label{RSW:pressure}
p_k[h]=\rho_k\eta_{k+1}[h]+\sum\limits_{l=1}^k\rho_lh_l=\rho_k\eta_k[h]+\sum\limits_{l=1}^{k-1}\rho_lh_l,
\end{equation}
where $\eta_k$ is the layer coordinates defined by $\eta_{L+1}=b$ with $b$ being  the bathymetry, and for $k=1,\dots,L$, $\eta_{k}[h]=b+\sum\limits_{l=k}^Lh_l$.
Following the constructions in \cite{pieper2019exponential,stewart2016energy}, the multi-layer Hamiltonian $\mathcal{H}$ is defined as
\begin{equation}
\mathcal{H}[V]=\sum\limits_{l=1}^k\rho_k\int\limits_{\Omega}\lb h_kK[\bu_k]+gh_k(\eta_{k+1}[h]+h_k/2)\rb\,{\rm d}{\bf x},
\end{equation}
and the skew-symmetric operator $\mathcal{J}[V]$ is
\begin{equation}
\mathcal{J}[V]\coloneqq\text{diag}_{k=1,\dots,L}\frac{1}{\rho_k}\mathcal{J}[h_k,\bu_k].
\end{equation}

\subsubsection{Jacobian Matrix}\label{Jacobian:sec}  
For the special ETD-Euler method (i.e. the exponential Rosenbrock-Euler) to be discussed in Section \ref{sect:etd}, one needs to evaluate the Jacobian matrix of the equation (\ref{RSW:eq}). Hence, we present the Jacobian matrix $J_k(h,\bu)$ here, that is,
\begin{equation}
J_k(h,\bu)=
\begin{bmatrix}
-\nabla\cdot\lb \bullet\bu_k\rb&-\nabla\cdot\lb h_k\bullet\rb\\
J_{2,1}	& J_{2,2}
\end{bmatrix},
\end{equation}
where the two entries $J_{2,1}$ and $J_{2,2}$ are defined by  
\begin{eqnarray*}
&&J_{2,1}=-\nabla\lb g/\rho_k\frac{\partial P_k}{\partial h}\rb-\frac{\partial q[h_k, \bu_k]}{\partial h}\hat{\bk}\times(h_k\bu_k)-q[h_k,\bu_k]\hat{\bk}\times(\bullet \bu_k),\\
&&J_{2,2}=-\nabla \bu_k-\frac{\partial q[h_k,\bu_k]}{\partial \bu}\hat{\bk}\times(h_k\bu_k)-q[h_k,\bu_k]\hat{\bk}\times(h_k\bullet), 
\end{eqnarray*}
and $\bullet$ is the place to fill in the elements when the Jocobian matrix works on a vector. 
The derivatives with respect to layer i's thickness and normal velocity are respectively
\begin{align*}
\frac{\partial P_k}{\partial h_i}&=\left\{
\begin{array}{lr}
\rho_k,& \text{if }i\geq k,\\
\rho_i,& \text{if }i<k, 
\end{array}
\right.\\
\frac{\partial q[h_k, \bu_k]}{\partial h_i}&=-\frac{\hat{\bk}\cdot\nabla\times\bu_k+f}{h_k^2}\delta_{i,k},\\
\frac{\partial q[h_k, \bu_k]}{\partial \bu_i}&=\frac{\hat{\bk}\cdot\nabla\times\bullet}{h_k}\delta_{i,k}.  
\end{align*}

Notice that evaluations of 
$-\nabla\cdot\lb \bullet\bu_k\rb$, $-\nabla\cdot\lb h_k\bullet\rb$ and $J_{2,2}$ only involve quantities from layer $k$ and, for calculating $J_{2,1}$, only the term $\frac{\partial P_k}{\partial h}$ uses the thickness and densities from the other layers. Therefore, we can further split the Jacobian matrix as the sum of two terms: 
\begin{equation}\label{Jac:split}
J_k(h,\bu)=J_k^P(h,\bu)+J_k^R(h,\bu),
\end{equation}
where 
$$J_k^P(h,\bu)=\begin{bmatrix}
0&0\\
-\nabla\lb g/\rho_k\frac{\partial P_k}{\partial h}\rb& 0
\end{bmatrix}
\text{ and } J_k^R = J_k(h,\bu)-J_k^P(h,\bu).
$$ 
Such splitting significantly simplifies the coding effort and improves the efficiency in the construction of the Jacobian matrices in our parallel implementation. 

\subsection{Multi-layer primitive equations}
	We consider the following multi-layer primitive equations with the z-level vertical coordinate: for $k=1,\dots,L,$
\begin{itemize}
	\item Thickness equation:
	\begin{equation}
	\frac{\partial h_k}{\partial t}+\nabla\cdot(h_k\bu_k)+\frac{\partial}{\partial z}(h_k w_k)=0.\label{layerthickness}
	\end{equation}
	\item Momentum equation:
	\begin{equation}
	\frac{\partial \bu_k}{\partial t}+\frac{1}{2}\nabla |\bu_k|^2+(\bk\cdot\nabla\times\bu_k)\bu_k^{\perp}+f\bu_k^{\perp}+w_k\frac{\partial\bu_k}{\partial z}=-\frac{1}{\rho_0}\nabla p_k+\nu_h\nabla^2\bu_k+\frac{\partial}{\partial z}(\nu_v\frac{\partial\bu_k}{\partial z}).\label{layer:momentum}
	\end{equation}
	\item Tracer equations:
	\begin{equation}
	\frac{\partial h_k\varphi_k}{\partial t}+\nabla\cdot(h_k\varphi_k\bu_k)+\frac{\partial}{\partial z}(h_k\varphi_k w_k)=\nabla\cdot(h_k\kappa_h\nabla\varphi_k)+h_k\frac{\partial}{\partial z}(\kappa_v\frac{\partial \varphi_k}{\partial z}).\label{tracer}
	\end{equation}
	\item Hydrostatic condition:
	\begin{equation}
	p_k=p_k^s(x,y)+\int_z^{z_k^s}\rho g\,{\rm d}z'.
	\end{equation}
	\item Equation of state:
	\begin{equation}
	\rho_k=f_{\text{eos}}(\Theta,S,p).
	\end{equation}
\end{itemize}
The variable definitions are listed in Table \ref{variable:def}.
\begin{table}[ht!]
	\centering
	\begin{tabular}{cl}
		\hline 
		Variables& Definition\\
		\hline
		$h_k$& Layer thickness\\
		$\bu_k$, $w_k$& Horizontal and vertical velocity\\
		$p_k$, $p_k^s(x,y)$& Pressure and the top pressure of layer $k$\\
		$\Theta$& Potential temperature\\
		$S$& Salinity\\
		$\rho_k$, $\rho_0$& Density and referential density\\
		$\varphi_k$ & Generic tracer ($\Theta$ or $S$)\\
		$z$, $z_k^s$& Vertical coordinate and the z-location of layer $k$'s top boundary location\\
		$\nu_h$, $\nu_v$& Viscosity\\
		$\kappa_h$, $\kappa_v$& Tracer diffusion\\
		\hline 
	\end{tabular}
	\caption{List of the variable definitions.}\label{variable:def}
\end{table}

As mentioned in the introduction, the primitive equations involve the barotropic mode for the dynamics and require a small time-stepping size for stable simulations with explicit time stepping schemes. To design an efficient numerical scheme, transformed dynamics equations are derived to treat the barotropic (fast) mode and baroclinic (slow) modes separately. Define the barotropic velocity $\overline{\bu}$ and baroclinic velocity $\bu_k'$ as
\begin{equation}
\overline{\bu}=\sum_{k=1}^{L}h_k\bu_k\Big/\sum_{k=1}^{L}h_k \,\text{  and  }\,
\bu_k'=\bu_k-\overline{\bu},\qquad k=1,\dots, L.
\end{equation}
Let $\zeta= \sum_{k=1}^{L}h_k+b-H$ 
be the sea surface height (SSH) with $b$ being the bottom elevation above a reference level and $H$  the height of the reference surface, and let $\Delta z_1$ be the top layer's original thickness. 
To derive the equation for $\zeta$, we consider the velocity $\lb u, v, w\rb$ where $ u, v$ are z-independent, then integrate the continuity equation over the entire fluid depth yields
\begin{equation}\label{inte:continuity}
\begin{split}
0=\int_{b}^{b+\sum_{k=1}^{L}h_k}\lb \frac{\partial u}{\partial x}+\frac{\partial v}{\partial y}+\frac{\partial w}{\partial z}\rb dz
=\lb \frac{\partial u}{\partial x}+\frac{\partial v}{\partial y}\rb\int_{b}^{b+\sum_{k=1}^{L}h_k}dz+w|_{b}^{b+\sum_{k=1}^{L}h_k}.
\end{split}
\end{equation}
The kinematic conditions \cite[Chapter 7]{cushman2011introduction} of the geophysical flow give the following two boundary conditions
\begin{equation*}
\begin{split}
w(z=b+{\textstyle\sum_{k=1}^{L}h_k})=\frac{\partial \zeta}{\partial t}+u\frac{\partial \zeta}{\partial x}+v\frac{\partial \zeta}{\partial y}, \quad
w(z=b)=u\frac{\partial b}{\partial x}+v\frac{\partial b}{\partial y}.
\end{split}
\end{equation*}
Plugging the above two equations into (\ref{inte:continuity}) gives
\begin{equation}\label{btr:ssh:eq}
\frac{\partial \zeta}{\partial t}=-\frac{\partial (u\sum_{k=1}^{L}h_k)}{\partial x}-\frac{\partial (v\sum_{k=1}^{L}h_k)}{\partial y}.
\end{equation}
For more details about $\zeta$, the readers are referred to \cite[Chapter 7]{cushman2011introduction}. 
By calculating the the layer-thickness-weighted average of (\ref{layer:momentum}) and combining (\ref{btr:ssh:eq}), the barotropic thickness and momentum equations are then given by
\begin{empheq}[left=\empheqlbrace]{align} 
\frac{\partial \zeta}{\partial t}&=-\nabla\cdot\lb\overline{\bu}\sum_{k=1}^{L}h_k\rb,\label{fast:ssh}\\
\frac{\partial\overline{\bu}}{\partial t}&=-f\overline{\bu}^{\perp}-g\nabla\zeta+\overline{\bm G}(\bu_k, \bu_k'),\label{fast:momentum}
\end{empheq}
where $\overline{\bm G}(\bu_k, \bu_k')$ includes all remaining terms in the barotropic equation. Subtracting (\ref{fast:momentum}) from (\ref{layer:momentum}) yields the baroclinic momentum equation: for $k=1,\dots, L$, 
\begin{equation}\label{slow}
\frac{\partial \bu_k'}{\partial t}=-f{\bu_k'}^{\perp}+\bT(\bu_k,w_k,p_k)+g\nabla\zeta-\overline{\bm G}(\bu_k, \bu_k'), 
\end{equation}
where 
$$\bT(\bu_k,\ w_k,p_k)=-\frac{1}{2}\nabla |\bu_k|^2-(\bk\cdot\nabla\times\bu_k)\bu_k^{\perp}-\ w_k\frac{\partial \bu_k}{\partial z}-\frac{1}{\rho_0}\nabla p_k+\nu_h\nabla^2\bu_k+\frac{\partial}{\partial z}(\nu_v\frac{\partial\bu_k}{\partial z}).$$

\section{The TRiSK schemes for spatial discretizations}\label{sect:trisk}
The multi-layer rotating SWE (\ref{RSW:eq}) and the barotropic and baroclinic equations (\ref{fast:ssh}-\ref{slow}) will be discretized by the mimetic TRiSK scheme \cite{thuburn2009numerical, ringler2010unified, thuburn2012framework} as done in {\em MPAS-Ocean} to ensure the properties of the continuous system, such as energy and mass conservations on unstructured, locally orthogonal meshes. 

\begin{figure}[ht!]
\centerline{
	\includegraphics[height=2.4in]{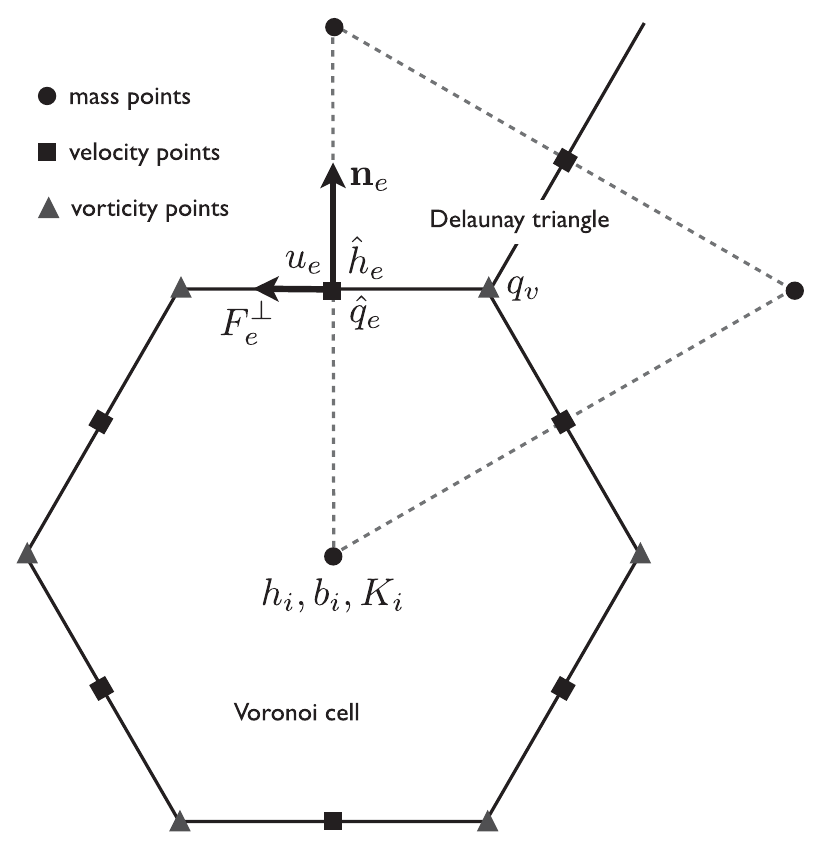}\hspace{0.5cm}
	\includegraphics[height=2.3in]{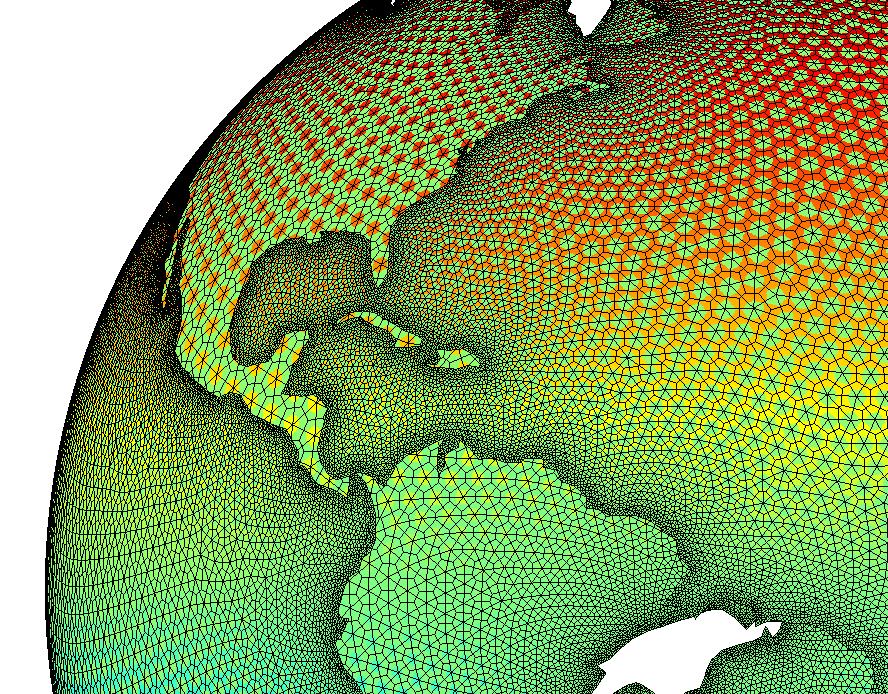}}
	\caption{An illustration of spherical centroidal Voronoi tessellation and its dual Delaunay triangulation for ocean modeling.}\label{scvt-Delunay}
\end{figure}

The TRiSK scheme utilizes the staggered C-grid, which is comprised of spherical centroidal Voronoi tessellation (SCVT) as the primal grid and its corresponding Delaunay triangulation as the dual grid, see Figure \ref{scvt-Delunay} for illustration. The layer thickness $h_k$, the vertical velocity $w$, the pressure $p$ and the tracer $\varphi$ are defined on the cell's center $\bx_i$, the horizontal velocity $\bu_k$ is defined at the midpoint $\bx_e$ of the edge, and the vorticity is located at the center $\bx_v$ of the dual grid's cell. We list all the notations for the mesh information from \cite{ringler2010unified} in Tables \ref{elements}-\ref{nortang}, and the corresponding differential operators in Table \ref{operator}.

\begin{table}[ht!]
	\centering
	\begin{tabular}{ll}
		\hline \hline
		Notations& Definition\\
		\hline
		$\bx_i$& Centers of primal mesh cells\\
		$\bx_v$& Centers of dual mesh cells\\
		$\bx_e$& The intersection point between the primal and dual edges\\
		$P_i$& Primal cells corresponding to $\bx_i$\\
		$D_v$& Dual cells corresponding to $\bx_v$\\
		 & \\
		$l_e$& Length of the primal edge $e$\\
		$d_e$& Length of the dual edge intersecting $e$\\
		$A_i$& Area of the primal cell $P_i$\\
		$A_{v}^{d}$& Area of the dual cell $D_v$\\
		$A_{e}^{e}$& Area associated with the primal edge $e$: $A_{e}^{e}=1/2l_ed_e$\\
		\hline \hline
	\end{tabular}
\caption{List of the mesh elements and quantities.}\label{elements}
\end{table}

\begin{table}[ht!]
	\centering
	\begin{tabular}{ll}
		\hline \hline
		Notations& Definition\\
		\hline
		$e\in EC(i)$& Set of edges of the primal cell $P_i$\\
		$i\in CE(e)$& Two primal cells either side of primal edge $e$\\
		$e\in EV(v)$& Set of primal edges sharing vertex $v$\\
		$i\in CV(v)$& Set of primal cells having $v$ as their vertex\\
		$v\in VE(e)$& Two endpoints of primal edge $e$\\
		$e'\in ECP(e)$& Set of primal edges nearby $e$\\
		\hline \hline
	\end{tabular}
\caption{List of the mesh connectivity.}\label{connect}
\end{table}

\begin{table}[ht!]
	\centering
	\begin{tabular}{ll}
		\hline\hline
		Notations& Definition\\
		\midrule
		$\bn_e$& The unit vector at $\bx_e$ normal to $e$ in the direction corresponding to positive $u_e$\\
$\bm{t}_e$& The unit tangential vector at $\bx_e$: $\bm t_e = \bm k \times \bn_e$ \\
$n_{e,i}$& Normal indicator function\\
		&   $n_{e,i}=\begin{cases}
		1 \quad \text{if }\bn_e \text{ is an outward normal of }P_i,     \\
		-1 \quad \text{ otherwise.}
		\end{cases}$\\
		$t_{e,v}$ & Tangential indicator function \\
		&   $t_{e,v}=\begin{cases}
		1 \quad \text{if }v \text{ is an outward normal of }\bm k\times\bn_e,    \\
		-1 \quad \text{ otherwise.}
		\end{cases}$\\
		&i.e., counterclockwise circulation about vertex $v$ contributes positively to the vorticity at $v$.\\
		\hline\hline
	\end{tabular}
\caption{List of the normal and tangential vectors.}\label{nortang}
\end{table}

\begin{table}[ht!]
\begin{center}
	\begin{tabular}{l r c l}
		\hline\hline
		Divergence : &$({\nabla\cdot}_{E\to I} \bm{y})_e$ & $=$ &$(1/A_i)\sum_{e\in EC(i)} n_{e,i} l_e \bm{y}_e$\\ 
		Gradient :& $({\nabla}_{I\to E} \bm{y})_i$ & $=$ & $(1/d_e)\sum_{i\in CE(e)} - n_{e,i} \bm{y}_i$\\ 
		Curl: &$({(\hat{\bm k}\cdot\nabla\times)}_{E\to V}\bm{y})_v$ & $=$ & $({1}/{A_v})\sum_{e\in EV(v)} t_{e,v} d_e \bm{y}_e $\\ 
		Perpendicular Gradient: &$({\nabla^{\perp}}_{V\to E} \bm{y})_e$ & $=$ & $({1}/{l_e})\sum_{v\in VE(e)}t_{e,v}\bm{y}_v$\\ 
		Perpendicular Flux: &$({\hat{\bm k}\times}_{E\to E} \bm{y})_e$ & $=$ & $({1}/{d_e})\sum_{e'\in ECP(e)}w_{e,e'}l_e\bm{y}_{e'}$\\ 
		Cell to Vertex interpolation: & $\{\bm{y}\}_{V,v}$ & $=$ & $({1}/{A_v})\sum_{i\in CV(v)}R_{i,v}A_i\bm{y}_i$\\ 
		Vertex to Edge interpolation: &$\{\bm{y}\}_{E,e}$ & $=$ & $\sum_{v\in VE(e)}\bm{y}_v/2$\\ 
		Edge to Cell interpolation: &$\{\bm{y}\}_{I,i}$ & $=$ & $({1}/{A_i})\sum_{e\in EC(i)}\bm{y}_eA_e/2$\\ 
		Cell to Edge interpolation: &$\{\bm{y}\}_{E,e}$ & $=$ & $\sum_{i\in CE(e)} \bm{y}_i/2$\\
		\hline\hline
	\end{tabular}
\end{center}
	\caption{Summary of the discrete operators given concretely in terms of
	geometrical quantities \cite{ringler2010unified}.}\label{operator}
\label{op_tab}
\end{table}
The discretization of the multi-layer SWE and primitive equations in TRiSK scheme are given as follow:
\begin{itemize}
	\item Multi-layer shallow water equations
	\begin{equation}\label{semisys}
	\left \{ \begin{array}{ll}
	& \frac{\partial h_k}{\partial t} =-\nabla_{E\rightarrow I} \cdot \lb\{h_k\}_E\ast\bu_k\rb, \vspace{0.2cm}\\
	& \frac{\partial \bu_k}{\partial t} =Q[h_k,\bu_k](\{h_k\}_E\ast\bu_k)-\nabla_{I\rightarrow E}\lb K\left[\left\{\bu_k\right\}_I\right]+(g/\rho_k)p_k[\{h\}_E]\rb + G_k( h,\bu) ,
	\end{array} \right .
	\end{equation} 
	where $\ast$ denotes the point- or component-wise product. The term $Q[h_k,\bu_k](\cdot)$ is an operator and varies accordingly with the time-stepping methods. Let
	$$
	q[h_k,\bu_k]=\lb{(\hat{\bm k}\cdot\nabla\times)}_{E\to V}\bu_k+f\rb/\{h_k\}_V,
	$$
	then we define
	\begin{eqnarray*}
	&&\hspace{-1cm}Q[h_k,\bu_k](\bm y)=\nonumber\\
	&&\hspace{-1cm}\qquad \left\{\begin{array}{ll}
	-\left\{q[h_k,\bu_k]\right\}_E\ast\lb {\hat{\bm k}\times}_{E\to E} \bm{y}\rb,& \text{ in ETD-Euler,}\\
	\frac{1}{2}\lb \left\{q[h_k,\bu_k]\right\}_E\ast\lb {\hat{\bm k}\times}_{E\to E} \bm{y}\rb+{\hat{\bm k}\times}_{E\to E}\left\{q[h_k,\bu_k]\right\}_E\ast \bm{y}\rb,& \text{ in ETD-Wave.}
	\end{array}
	\right.
	\end{eqnarray*}
	The second form comes from the skew-symmetry of ${\hat{\bm k}\times}_{E\to E}$.
	\item Multi-layer primitive equations
	\begin{equation}\label{prim:trisk}
	\left \{\begin{array}{ll}
	&\frac{\partial h_k}{\partial t}+\nabla_{E\rightarrow I} \cdot \lb\{h_k\}_E\ast\bu_k\rb+\frac{\partial}{\partial z}(h_k\ w_k)=0,\\
	&\\
	&\frac{\partial \bu_k}{\partial t}+\nabla_{I\rightarrow E}K\left[\left\{\bu_k\right\}_I\right]
	+\lb\left\{{(\hat{\bm k}\cdot\nabla\times)}_{E\to V}\bu_k\right\}_E+f\rb \ast \lb{\hat{\bm k}\times}_{E\to E}\bu_k\rb
	+\ w_k\frac{\partial\bu_k}{\partial z}\\
	&\quad=-\frac{1}{\rho_0}\nabla_{I\rightarrow E} p_k
	+\nabla_{I\rightarrow E}(\nabla_{E\rightarrow I}\cdot\bu_k)+{\nabla^{\perp}}_{V\to E}{(\hat{\bm k}\cdot\nabla\times)}_{E\to V}\bu_k
	+\frac{\partial}{\partial z}(\nu_v\frac{\partial\bu_k}{\partial z}),\\
	&\\
	&\frac{\partial h_k\varphi_k}{\partial t}+\nabla_{E\rightarrow I}\cdot(\{{h_k}\varphi_k\}_E\ast\bu_k)+\frac{\partial}{\partial z}({h_k}\varphi_k\ w_k)=\nabla_{E\rightarrow I}\cdot(\kappa_h\{h_k\}_E\ast\nabla_{I\rightarrow E}\varphi_k)\\
	&\qquad+h_k\frac{\partial}{\partial z}(\kappa_v\frac{\partial \varphi_k}{\partial z}),
	\end{array}
	\right.
	\end{equation}
	where we applied $\nabla^2\bu=\nabla(\nabla\cdot\bu)+\bm k\times \nabla(\bm k\cdot\nabla\times\bu)$. For the vertical derivative discretization, we will refer the methods from \cite{ringler2013multi} that are shown below. First of all, let $\phi$ be a generic variable and set
\begin{align*}
\startsubequation\subequationlabel{eq:3}
\overline{\lb \phi_{:}^{t}\rb}_{k}^{m}&=(\phi_{k}^{t}+\phi_{k+1}^{t})/2,\\
\overline{\lb \phi_{:}^{m}\rb}_{k}^{t}&=(\phi_{k-1}^{m}+\phi_{k}^{m})/2,\\
\delta z_{k}^{m}(\phi_{:}^{t})&=\frac{\phi_{k}^{t}-\phi_{k+1}^{t}}{h_k},\\
\delta z_{k}^{t}(\phi_{:}^{m})&=\frac{\phi_{k-1}^{m}-\phi_{k}^{m}}{\overline{(h)}_{k}^{t}},
\end{align*}
then we can discretize the vertical derivatives as
\begin{align*}
\frac{\partial}8{\partial z}(h_k\ w_k)&\approx \omega_{k}^{t}-\omega_{k+1}^{t},\\
\ w_k\frac{\partial\bu_k}{\partial z}&\approx \overline{\lb \omega_{:}^{t}\delta z^t(\bu_{:})\rb}_{k}^{m},\\
\frac{\partial}{\partial z}(\nu_v\frac{\partial\bu_k}{\partial z})&\approx \delta z_{k}^{m}(\nu_v \delta z^t(\bu_{:})),\ \frac{\partial}{\partial z}(\kappa_v\frac{\partial \varphi_k}{\partial z})\approx \delta z_{k}^{m}(\nu_v \delta z^t(\varphi_{:}))\\
\frac{\partial}{\partial z}({h_k}\varphi_k\ w_k)&\approx \overline{\lb \varphi_{:}^{m}\rb}_{k}^{t}\omega_{k}^{t}-\overline{\lb \varphi_{:}^{m}\rb}_{k+1}^{t}\omega_{k+1}^{t}.
\end{align*}
\end{itemize}

\section{Exponential time differencing methods and parallel implementations}\label{sect:etd}
In this section, we first briefly go over and discuss the classic ETD method \cite{hochbruck2009exponential} and the recently derived ETD-wave and Barotropic ETD-wave methods in \cite{pieper2019exponential} for the multi-layer SWEs. Then, we propose a new two-level ETD method to solve the multi-layer  primitive equations in the barotropic/baroclinic splitting fashion. Finally we discuss their parallel implementation through domain decomposition.

\subsection{Classic ETD schemes}
Consider the following nonlinear parabolic equation on the time interval $[t_n,t_n+\tau]$:
\begin{equation}\label{lin:non:split}
\partial_tV=F[V]=A_nV+N[V],
\end{equation}
where $V$ is the unknown variable, $F[V]$ is the flux, $A_n$ is a time-independent linear operator, and $N[V]$ is a nonlinear term. 
By using the integrating factor $e^{-t\Delta A_n}$, (\ref{lin:non:split}) can be advanced to a future time $t_n+\Delta t$ provided $V_n$, the value of $V$ at $t_n$:
\begin{equation}\label{lin:non:split:sol}
\begin{split}
V(t_n+\Delta t)&=e^{\Delta t A_n}V_n+\int_{t_n}^{t_n+\Delta t}e^{(t_n+\Delta t-\tau)A_n}N(V(\tau))\,{\rm d}\tau\\
&=V_n+\Delta t\varphi_1(\Delta t A_n)A_nV_n+\Delta t\int_{0}^{1}e^{\Delta t(1-\theta)A_n}N(V(t_n+\theta \Delta t))\,{\rm d}\theta.
\end{split}
\end{equation}
If we replace $N(V(t_n+\theta \Delta t))$ with a polynomial approximation in time
\begin{equation}
N(V(t_n+\theta \Delta t))\approx \sum_{s=2}^{S}\frac{\theta^{s-1}}{(s-1)!}b_{n,s},
\end{equation}
where $b_{n,s}$ approximates the derivatives $\frac{d^{s-1}N(V(t_n+\theta \Delta t))}{d \theta^{s-1}}\big|_{\theta=0}$, then $V(t_{n+1})$ can be approximated with $S$ internal stages as 
\begin{equation*}
V(t_{n+1})\approx V_{n+1}=V_n+\Delta t\Big(\varphi_1(\Delta tA_n)F[V_n]+\sum_{s=2}^S\varphi_s(\Delta tA_n)b_{n,s}\Big),
\end{equation*}
where the $\varphi-$functions are defined by
\begin{equation}
\varphi_s(z)=\int_0^1\exp((1-\sigma)z)\frac{\sigma^{s-1}}{(s-1)!}\,{\rm d}\sigma=\sum_{k=0}^{\infty}\frac{z^k}{(k+s)!}.
\end{equation}
We call the method with $S=1$, ETD-Euler method, that is
\begin{equation}\label{etd:euler}
V_{n+1}=V_n+\Delta t\varphi_1(\Delta tA_n)F[V_n].
\end{equation}
If $A_n=\frac{\partial F[V]}{\partial V}\big|_{t=t_n}$, then (\ref{etd:euler}) is also specially called 
{\it exponential Rosenbrock-Euler}  method, which is of second order accuracy in time for  autonomous systems. Let us focus on the space-discrete case, i.e.
\eqref{lin:non:split} is a ODE system where $V$ is a vector and $A_n$ is a matrix.
To compute $\varphi_1(\Delta tA_n)F[V_n]$, or more generally $\varphi_s(A_n)b$, we will use Krylov subspace methods \cite{gs92,saad92}. The essential idea is to approximate the matrix $A_n$ with a matrix $H_M$ of smaller size but preserves as much information of $A_n$ as possible. Usually, it can be done in two steps.
\begin{itemize} 
\item Firstly, one constructs the orthonormal basis $\bV_M$ of the Krylov subspace 
\begin{equation}
K_M(A_n,b)=\text{span}\{b,A_nb,\cdots,A_n^{M-1}b\}
\end{equation}
by the Arnoldi process. It also generates an unreduced upper Hessenberg matrix $\boldmath{H}_M$, satisfying the recurrence formula
$$ A_n\bV_M=\bV_M\boldmath{H}_M+\bm h_{M+1,M}\bv_{M+1}\be_M^T,\ \bV_M=(v_1, v_2, \cdots, v_M).$$
Hence, 
\begin{eqnarray}\label{varphi:afterKry}
\varphi_s(\Delta tA_n)b\approx \bV_M\varphi_s(\bV_M^T\Delta tA_n\bV_M)\bV_M^T b=\| b\|_0\bV_M\varphi_s(\Delta t\boldmath{H}_M)\be_1, 
\end{eqnarray}
where $\|b\|_0=\sqrt{b^Tb}$. 
If the matrix $A_n$ has some (skew-)symmetric properties, then the Arnoldi process can be replaced by the (skew-)Lanczos process to reduce the Gram-Schmidt process and obtain a tridiagonal matrix $H_M$.
\item Secondly, one instead computes $\varphi_s(\Delta t\boldmath{H}_M)e_1$ and then $\| b\|_0\bV_M\varphi_s(\Delta t\boldmath{H}_M)\be_1$
as an approximation of $\varphi_s(A_n)b$ due to \eqref{varphi:afterKry}. For that, many methods such as Pad\'e approximation or scaling-doubling method can be applied \cite{moler2003nineteen,higham2005scaling,sidje1998expokit}. In our numerical tests in Section \ref{sect:parallel}, we will use the function ``dgpadm" from the software package Expokit \cite{sidje1998expokit} which uses the scaling-doubling method.

\end{itemize}
\subsection{ETD-wave method for multi-layer SWEs}
The linearization of the Hamiltonian formalism (\ref{Hamilton:single}) of the multi-layer SWEs around the reference state $V^{\text{ref}}=\lb h^{\text{ref}}, \bu^{\text{ref}}\rb$ is given by
\begin{equation}\label{wt}
\partial_tW=\mathcal{J}'[V^{\text{ref}};W]\delta \mathcal{H}[V^{\text{ref}}]+\mathcal{J}[V^{\text{ref}}]\delta^2\mathcal{H}[V^{\text{ref}}]W,\quad W(0)=W_0,
\end{equation}
where $V=V^{\text{ref}}+W$ and $W=\lb h_w, \bu_w\rb$ is the perturbation. Picking $\bu^{\text{ref}}=0$, then (\ref{wt}) is reduced to
\begin{equation}\label{wave:app}
\partial_tW=\mathcal{J}[V^{\text{ref}}]\delta^2\mathcal{H}[V^{\text{ref}}]W=AW,
\end{equation}
where $\mathcal{J}$ and $\delta^2\mathcal{H}$ have the following symmetry properties,
\begin{equation*}
\lb \bW, \mathcal{J}\bV\rb_{\mathcal{X}^L}=-\lb \mathcal{J}\bW, \bV\rb_{\mathcal{X}^L},\quad \lb \bW, \delta^2\mathcal{H}\bV\rb_{\mathcal{X}^L}=\lb \delta^2\mathcal{H}\bW, \bV\rb_{\mathcal{X}^L}.
\end{equation*}
In order to apply the symmetry property on the discrete level, we first specify some spaces for the discrete quantities. Let $\bX=\bX_{I}\times\bX_{E}\in R^{N_I}\times R^{N_E}$, where $N_I$ and $N_E$ are the numbers of cells and edges on each layer. 
We assume for each layer $h_k\in \bX_{I}$, $ \bu_k\in\bX_{E}$, and the whole system solution $(h,\bu)\in \bX^L=\bX_{I}^{L}\times \bX_{E}^{L}$. 
Define the mass matrix $\bm M_{\bX^L}$ containing L copies of the cell and edge areas on the diagonal, then
\begin{equation*}\label{skew:A}
\bm M_HA=-A^T\bm M_H, \quad\text{where }\bm M_H=\bm M_{\bX^L}\delta^2\mathcal{H}.
\end{equation*}

Based on this property, we can introduce the skew-Lancos process to calculate $\varphi_s(\Delta tA_n)b$ at each time step under the $\bm M_{\bX^L}-$norm, which is defined as
\begin{equation*}
\lb u, v\rb_{\bm M_{\bX^L}}=u^T\bm M_{\bX^L}v,
\end{equation*}
where $u, v$ are two vectors. The method, ETDSwave, is to pick ``S" internal stages, and use the approximation (\ref{wave:app}) with the skew-Lancos process to find the lower dimensional matrix $\bm H_M$, then calculate the matrix exponential function.

\subsection{Barotropic ETD-wave method for multi-layer SWEs}
The barotropic ETD-wave method for solving multi-layer SWEs approximates the whole system by using some fast modes.  
It introduces a reduced-layer space
$$\bX^{\widehat{L}}=\bX_{I}^{\widehat{L}}\times \bX_{E}^{\widehat{L}}, \quad \text{ for }1\leq \widehat{L}<<L,$$
and define the mapping $\Psi: \bX^{\widehat{L}}\rightarrow \bX^{L}$ with the structure
\[
\Psi=\begin{bmatrix}
\Psi_h&0\\
0&\Psi_u 
\end{bmatrix},
\]
where $\Psi_h$ and $\Psi_u$ are defined by
\begin{equation*}
\begin{split}
[\Psi_h\hat{h}]_i&=\sum_{j=1}^{\widehat{L}}\Psi_h^{j,i}\hat{h}_{k,i}, \forall\hat{h}\in \bX_{I}^{\widehat{L}}\\
[\Psi_u\hat{\bu}]_i&=\sum_{j=1}^{\widehat{L}}\Psi_{\bu}^{j,i}\hat{\bu}_{k,i}, \forall\hat{\bu}\in \bX_{E}^{\widehat{L}}. 
\end{split}
\end{equation*}
Here $\hat{h}_{k,i}$ and $\hat{\bu}_{k,i}$ are the $i$th components of $\hat{h}_k$ and $\hat{\bu}_k$, respectively. $\Psi_h^{j,i}$ and $\Psi_{\bu}^{j,i}$ correspond to the $j$th fastest vertical height mode \cite[Section 2.3]{pieper2019exponential}. Then we define the reduced Hamiltonian as
\begin{equation}
\hat{\mathcal{H}}^{\text{ref}}(\widehat{V})=\mathcal{H}(\Psi\widehat{V})=\frac{1}{2}\lb \Psi\widehat{V}, \delta^2\mathcal{H}\Psi\widehat{V}\rb_{\mathcal{X}^L}=\frac{1}{2}\lb \widehat{V}, \delta^2\hat{\mathcal{H}}\widehat{V}\rb_{\mathcal{X}^L},
\end{equation}
where $\delta^2\hat{\mathcal{H}}$ is the corresponding reduced Hamiltonian matrix and $ \delta^2\hat{\mathcal{H}}=\Psi^T\delta^2\mathcal{H}\Psi$. As we can see that the mapping $\Psi$ is singular, we will define its inverse mapping through the Moore-Penrose pseudoinverse
\begin{equation}
\Psi^{\dagger}: \bX^L\rightarrow \bX^{\widehat{L}}, \ \Psi^{\dagger}=\lb \delta^2\hat{\mathcal{H}}\rb^{-1}\Psi^T\delta^2\mathcal{H}.
\end{equation}
The following proposition shows $\Psi^{\dagger}$ optimally projects the full solution $V$ onto the reduced layer space $\bX^{\widehat{L}}$ under the energy norm $\|\cdot\|_{\delta^2H}$.
\begin{proposition}{\rm \cite[Proposition 5.1]{pieper2019exponential}}
	The restriction matrix $\Psi^{\dagger}$ gives the solution to the following minimization problem: for any $V\in \bX^L$, we have $ \Psi^{\dagger}V=\widehat{V}$ where
	\begin{equation}
	\widehat{V}=\argmin_{\widehat{V}\in \bX^{\widehat{L}}} \big\| V-\Psi\widehat{V}\big\|_{\delta^2H}=\argmin_{\widehat{V}\in \bX^{\widehat{L}}}\frac{1}{2}\lb V-\Psi\widehat{V}, \delta^2H(V-\Psi\widehat{V})\rb_{\bX^L}. 
	\end{equation}
\end{proposition}
We can define the orthogonal projection $P:\bX^{L}\rightarrow\bX^{L}$ as
\begin{equation*}
\bP=\Psi\Psi^{\dagger},\quad \text{ such that } \bP V=\Psi\widehat{V}.
\end{equation*}
Then, we define
\begin{equation*}
\begin{split}
A_p=\bP A\bP=\Psi \widehat{A}\Psi^{\dagger},
\quad\text{ where }\widehat{A}=\Psi^{\dagger}A\Psi.
\end{split}
\end{equation*}
Based on the above framework of the reduced layer model, we can reformulate $\varphi_s(A)$ with respect to the reduced matrix $\widehat{A}$.
\begin{proposition}{\rm \cite[Proposition 5.3]{pieper2019exponential}}
	Let $A_p=\bP A\bP$ and $\widehat{A}=\Psi^{\dagger}A\Psi$. Then, it holds for any $s\geq0$ that
	\begin{equation}
	\varphi_s(A_p)=\frac{1}{s!}(Id-\bP)+\Psi\varphi_s(\widehat{A})\Psi^{\dagger}. 
	\end{equation}
\end{proposition}

Thereby, we can compute $\varphi_s(\Delta t\widehat{A})\hat{b}_s$ instead of $\varphi_s(\Delta tA_p)b_s$ in the ETD method. Since the barotropic modes usually much faster than the remaining modes in the realistic global ocean simulations, so we can choose the fast barotropic mode and the obtained ETD method is referred as B-ETDSwave method in \cite{pieper2019exponential} with ``S" denoting the number of internal stages.

\subsection{Two-level ETD method for multi-layer primitive equations}\label{subsect:etdsp}
By following the framework of the two-level time-stepping method \cite{higdon2005two} currently used in MAPS-Ocean, we next propose a new two-level ETD-based method for solving the multi-layer primitive equations, in which the same time step size $\Delta t$ is able to be used in advancing both barotropic and baroclinic modes.  The following three stages will be repeated twice in a predictor-corrector way.
\begin{itemize}
	\item \textbf{Stage 1: Advance the baroclinic mode (3D) explicitly. }
	
	As suggested in \cite{higdon2005two}, we first ignore the barotropic forcing term $\overline{\bm G}$, and apply the forward Euler method to predict the baroclinic velocity as
	\begin{equation}\label{forward:1st}
	\tilde{\bu}_{k,n+1}'=\bu_{k,n}'+\Delta t\lb -f \bu_{k,n}^{\perp}+\bm T^u+g\nabla\zeta_n\rb. 
	\end{equation}
Due to the layer thickness average free property of the baroclinic velocity, that is, 
	$$\sum_{k=1}^{L}h_k{\bu}_{k,t}'\Big/\sum_{k=1}^{L}h_k=0,$$
 we can obtain 
	\begin{equation}\label{forward:G}
	\overline{\bm G}=\frac{1}{\Delta t}\sum_{k=1}^{L}h_k\tilde{\bu}_{k,n+1}'\Big/\sum_{k=1}^{L}h_k,
	\end{equation}
	and then correct the baroclinic velocity by
	\begin{equation}\label{forward:cor}
	\bu_{k,n+1}'=\tilde{\bu}_{k,n+1}'-\Delta t\overline{\bm G}.
	\end{equation}
	\item \textbf{Stage 2: Compute the barotropic mode (2D) by ETD method.}
	
	At this stage, we are going to solve (\ref{fast:ssh})-(\ref{fast:momentum}). The barotropic forcing term $\overline{\bm G}$ in (\ref{fast:momentum}) is obtained from Stage 1. These two equations can be put in the general form:
	\begin{equation}\label{barotropic:eqn}
	\frac{\partial V}{\partial t}=-\bm F(V)+\bm b,
	\end{equation}
	where $V=\lb \zeta, \overline\bu\rb^T$, $F(V)=\lb \nabla\cdot(\overline{\bu}\sum_{k=1}^{L}h_k), f\overline{\bu}^{\perp}+g\nabla\zeta\rb^T$, and $\bm b=\lb 0,\overline{\bm G}\rb^T$. The associated Jacobian matrix is
	\begin{equation}
	J_n\coloneqq\frac{\partial F(V)}{\partial V}\bigg|_{t=t_n}=
	\begin{bmatrix}
	-\nabla\cdot\lb \bullet\overline\bu_n\rb&-\nabla\cdot\lb \bullet\sum_{k=1}^{L}h_{n,k}\rb\\
	-g\nabla \bullet	& -f\bk\times \bullet
	\end{bmatrix}.
	\end{equation}
	According to (\ref{lin:non:split:sol}), given $V_n$, we can advance (\ref{barotropic:eqn}) as 
	\begin{equation}\label{etd:stage2}
	\left\{
	\begin{split}
	V_{n+1}&
	= V_n+\Delta t\varphi_1(-\Delta J_n)\lb -\bm F(V_n)+ \bm b_n\rb,\\
	V_{n+1/2}&=1/2(V_n+V_{n+1}),
	\end{split}
	\right.
	\end{equation}
where we choose the left endpoint rule for the integral. This method is referred  as ``ETD-SP1". 
As mentioned in \cite{mit2004}, it essentially subsamples the high-frequency barotropic motions and consequently alias high-frequency energy onto lower frequencies if we only advance the barotropic mode to $t_{n+1}$. There is one solution suggested in \cite{mit2004} to average over the baroclinic time step, and this is best done by integrating 
the barotropic equations forward over $2\Delta t$, then average with the solution at $t_n$. So we will consider the following approximation to calculate $\overline{\bu}_{n+1}$
	\begin{equation}\label{etd:stage2:sp2}
	\left\{
	\begin{split}
	V_{n+2}&\approx V_n+2\Delta t\varphi_1(-2\Delta t J_n)\lb -\bm F(V_n)+ \bm b_n\rb,\\
	V_{n+1}&=1/2(V_n+V_{n+2}).
	\end{split}
	\right.
	\end{equation}
This method is referred as ``ETD-SP2". Accordingly, we will refer the split-explicit methods currently adopted in MPAS-Ocean as ``SP1" or ``SP2" for advancing the barotropic mode to $\Delta t$ or $2\Delta t$.
	\item \textbf{Stage 3: Update thickness, tracers, density and pressure by forward-Euler scheme.}
	
Notice that the equations belong to the hyperbolic type of advection equations. As suggested in \cite{smolarkiewicz1998mpdata}, we better choose the transport velocity at the intermediate time level $t_{n+1/2}$. While the transport velocity is split into two parts, we approximate the baroclinic velocity at $t_{n+1/2}$  by averaging the velocities at the two consecutive time steps. However, we will consider the barotropic velocity at $t_{n+1}$ due to the sub-sampling issue mentioned in Stage 2.
\end{itemize}

\subsection{Domain decomposition and parallel implementations}\label{sect:parallel}
The major computational cost for the above ETD methods lies on  evaluations of matrix exponential and vector products. Even if the Krylov subspace methods are used, a large amount of matrix-vector products are still required in both Arnoldi and Lanczos processes. Therefore, parallel implementation of these ETD methods on large distributed systems mainly focuses on parallelizing these products efficiently. Due to its cross-platform portability and high performance, our code is developed with the message passing interface (MPI) \cite{gropp1999using}. 

\begin{figure}[ht!]
\centerline{
	\includegraphics[scale=0.38]{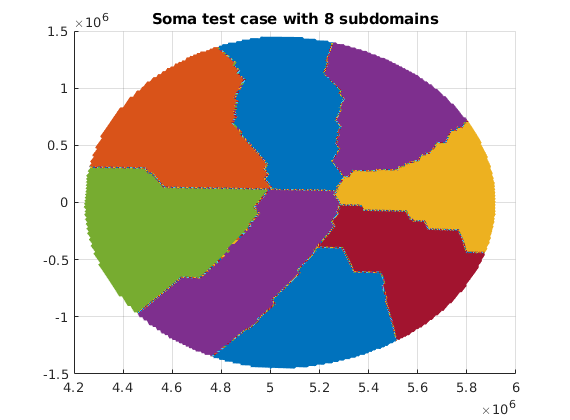}\hspace{-0.5cm}
	\includegraphics[scale=0.38]{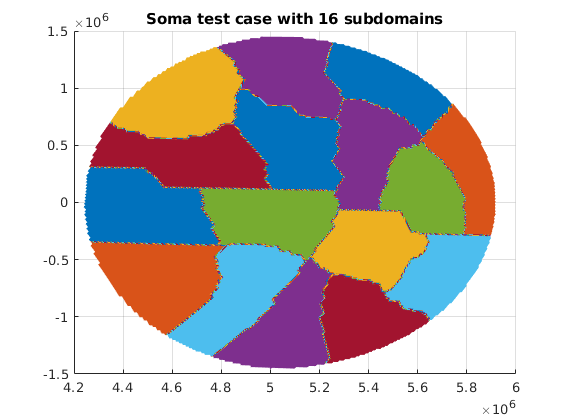}\hspace{-0.5cm}
	\includegraphics[scale=0.38]{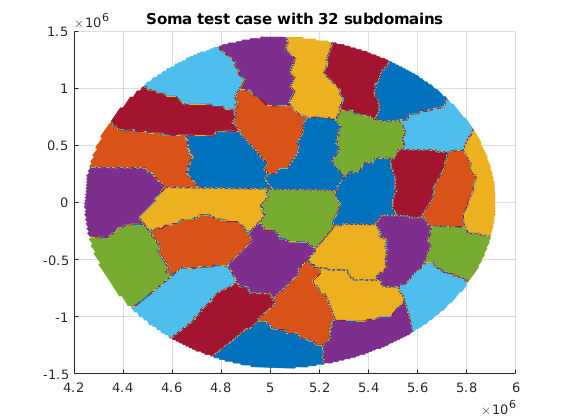}}
	\caption{Domain decomposition examples for the SOMA test cases.}\label{partition:metis}
\end{figure}

\begin{figure}[ht!]
\centerline{
	\includegraphics[width=2.3in,height=2.2in]{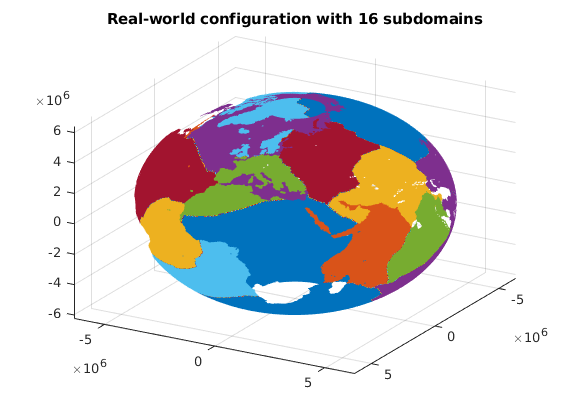}\hspace{-0.5cm}
	\includegraphics[width=2.3in,height=2.2in]{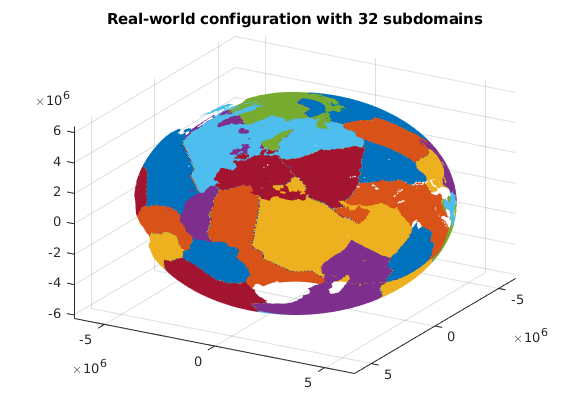}\hspace{-0.5cm}
	\includegraphics[width=2.3in,height=2.2in]{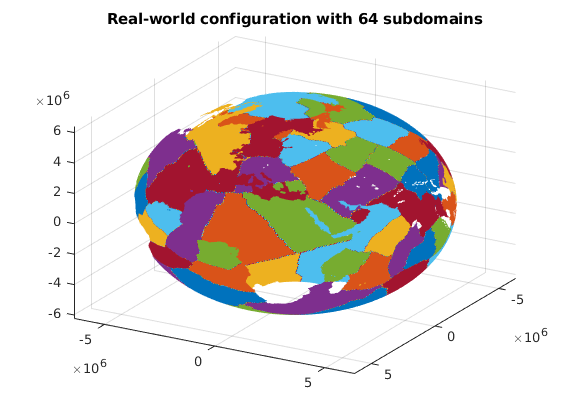}}
	\caption{Domain decomposition examples for the global ocean test case.}\label{partition:metis:realworld}
\end{figure}

For both layered SWE and primitive equation models, we first partition the target three-dimensional domain into several vertical layers, and each layer shares the same horizontal domain meshed by the SCVTs together with a dual Delaunay triangulation. 
To achieve high parallelization efficiency, we further decompose the horizontal domain into $N_p$ subdomains, where $N_p$ is the number of the processors. Six horizontal partitions generated by ``METIS" \cite{karypis1998fast} are shown in Figures \ref{partition:metis} and \ref{partition:metis:realworld}. To avoid the communication between processors due to the pressure, we assign all the vertical layers relating to the same sub-domain to the same processor. More specifically, suppose the mesh nodes are labeled as $\{N_{i,k}\;|\; i=1,\dots,M,\ \text{and }k=1,\dots,L\}$, where M is the number of nodes on the horizontal mesh, L is the number of layers. The indices $1,\dots,M$ will be separated into $N_p$ groups based on the domain decomposition, i.e., there are $N_p$ index sets: $\tilde{I}_1,\cdots, \tilde{I}_{N_p}$, such that $\tilde{I}_i\cap \tilde{I}_j=\emptyset, \text{ for }i\ne j.$ In order to share the information among processors, we need to assign some interface cells from their neighbor subdomains. In our   implementation, three outer layer cells will be used. This would lead to new index sets $I_1,\cdots,I_{N_p}$, such that $\tilde{I}_i\subset I_i$. Hence, the $i$th process is assigned with the indices $\{N_{i,k}\;|\; i\in I_i,\ \text{and }k=1,\dots,L\}$. 
We  use the sophisticated parallel package Epetra \cite{epetra-website} from Trilinos \cite{trilinos-website} to take care of the MPI data structure and inter-process communication.


\section{Numerical experiments}\label{sect:experiments}
In \cite{pieper2019exponential}, the authors tested ETD2wave and B-ETD2wave methods on multi-layer shallow water equations (\ref{RSW:eq}). Hence in this paper, we mainly focus on their parallel implementation. We choose the test cases from MPAS-Ocean Version 7.0. Without special notification, our numerical experiments are carried on a project-owned partition of the Cori at the National Energy Research Scientific Computing Center (NERSC). Cori is a Cray XC40 with a peak performance of about 30 petaflops, which is comprised of 2,388 Intel Xeon``Haswell" processor nodes and 9,688 Intel Xeon Phi ``Knight's Landing" (KNL) nodes. Our codes run on  "Haswell" processor nodes, where each node has two 16-core Intel Xeon ``Haswell" (E5-2698 v3, 2.3 GHz) processors and 128 GB DDR4 2133 MHz memory.

\subsection{The SOMA test case for the shallow water equations with  three layers}\label{soma:test}
\begin{figure}[!ht]
	\centering{
		\includegraphics[height=1.5in]{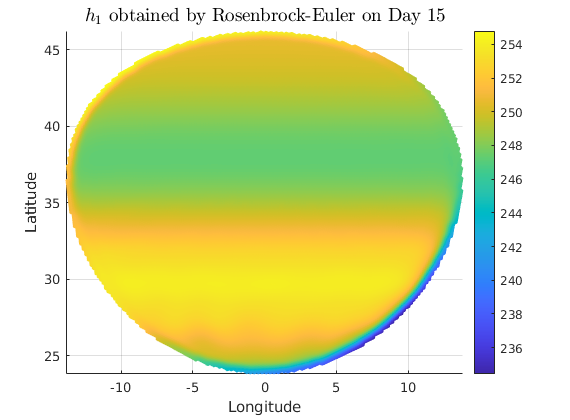}
		\includegraphics[height=1.5in]{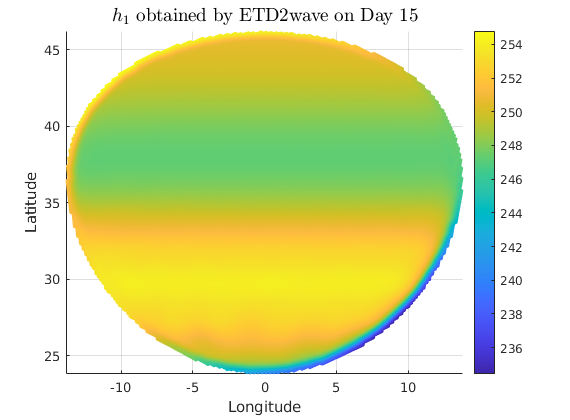}
		\includegraphics[height=1.5in]{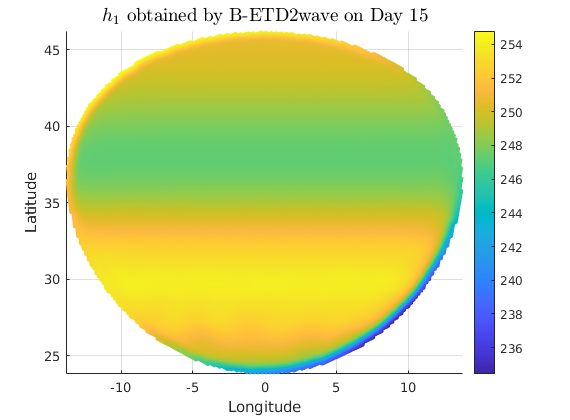}
	} 
	\centering{
		\includegraphics[height=1.5in]{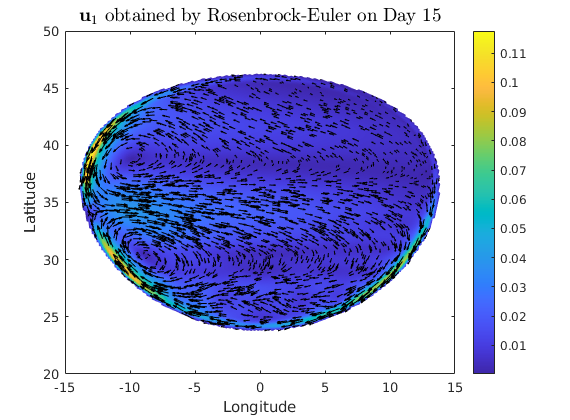}
		\includegraphics[height=1.5in]{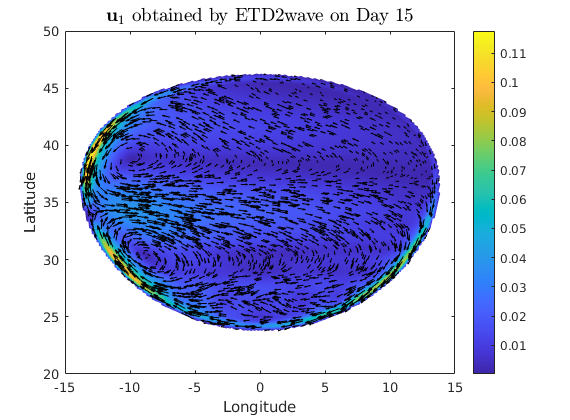}
		\includegraphics[height=1.5in]{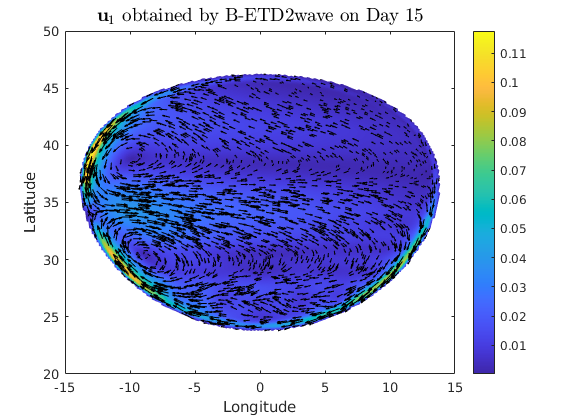}
	} 
	\caption{Simulated layer thickness (top row) and velocity (bottom row) ) of the first layer on Day 15 by ETD methods  for the SOMA test case. Latitude and Longitude are in degrees.}\label{15days:soma:thickness-velocity}
\end{figure}

\begin{figure}[!ht]
	\centering{
		\includegraphics[height=1.5in]{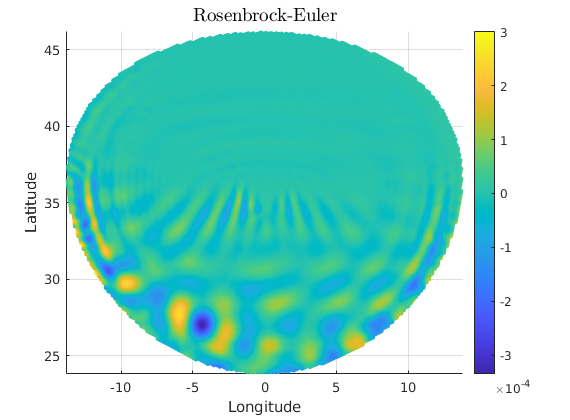}
		\includegraphics[height=1.5in]{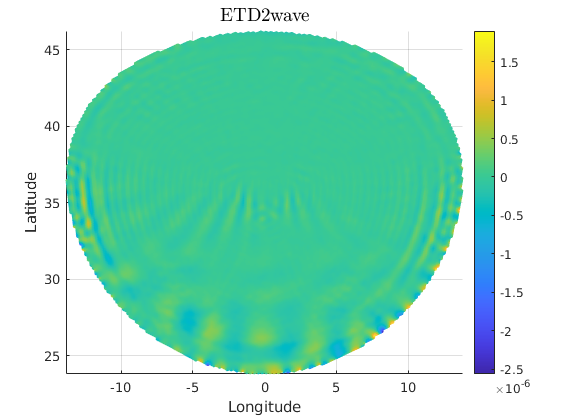}
		\includegraphics[height=1.5in]{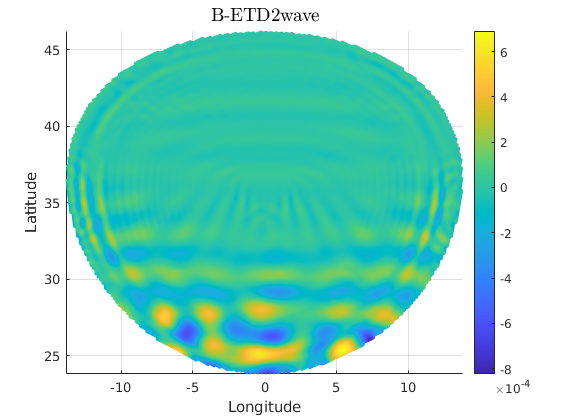}
	} 
	\centering{
		\includegraphics[height=1.5in]{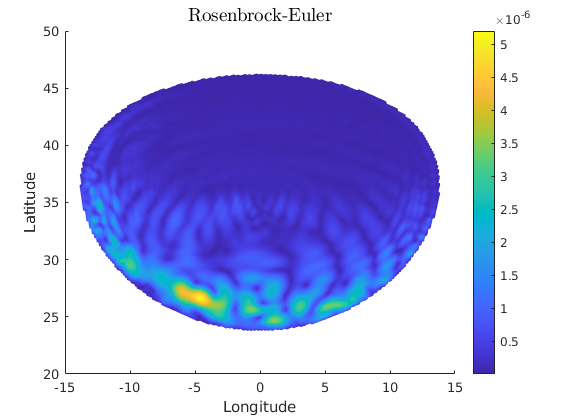}
		\includegraphics[height=1.5in]{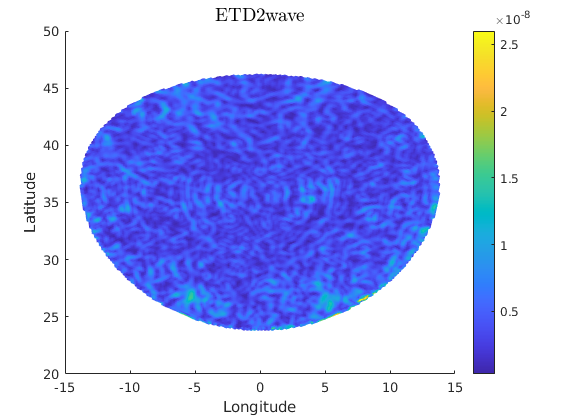}
		\includegraphics[height=1.5in]{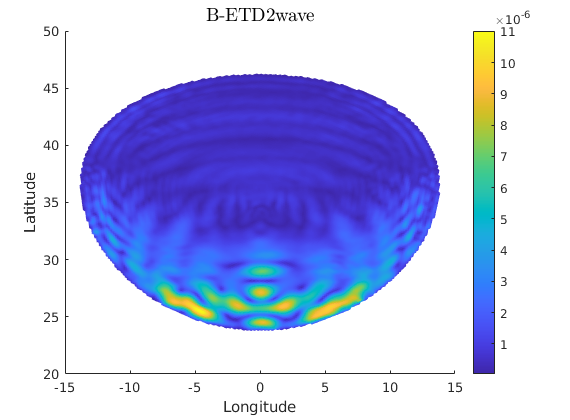}
	} 
	\caption{The differences in simulated layer thickness (top row) and velocity (bottom row) of the first layer on Day 15 between the ETD solutions and the RK4 solution  for the SOMA test case. Latitude and Longitude are in degrees.}\label{15days:soma:thickness-velocity:diff}
\end{figure}

In this test, the spatial domain is a circular basin centered at the point $\bx_c$ (latitude $\theta_c=35^{\circ}$ and longitude $\alpha_c=0^{\circ}$) with radius 1250 km, lying on the surface of the sphere of radius $R=6371.22$ km. The fluid depth in the basin varies from 2.5 km at the center to 100 m on the coastal shelf. The initial interfaces of the three-layer locations are at $\eta_1^0=0$ m, $\eta_2^0=-250$ m, and $\eta_3^0=-700$ m and the layer densities are $(\rho_1, \rho_2, \rho_3) = (1025, 1027, 1028)$ kg/m$^3$. Three SCVT meshes of different resolutions on each layer are used:
\begin{itemize}
	\item [i.]16 km resolution with 22,007 cells, 66,560 edges and 44,554 vertices;
	\item [ii.]8 km resolution with 88,056 cells, 265,245 edges and 177,190 vertices;
	\item [iii.]4 km resolution with 352,256 cells, 1,058,922 edges and 706,667 vertices.
\end{itemize}
The maximum dimension of Krylov subspaces used in ETD methods is set  to  be 25. 
{We run 15-day simulations by the three parallel ETD methods with the time step-size $\Delta t = 107$ s. In addition, we carry out the same simulation using the fourth-order Runge-Kutta (RK4) method with the small time step-size $\Delta t = 10.7$ s and  take its results  as the benchmark  solution for computing errors. The simulated  layer thickness $h_1$ and velocity $\bu_1$ of the first layer by parallel implementations of the ETD methods are shown in Figure \ref{15days:soma:thickness-velocity}.  The corresponding  differences between these solutions and the RK4 solution are shown  in Figure \ref{15days:soma:thickness-velocity:diff}, together with the quantitative results of relative $l_{\infty}$ errors  reported in Table \ref{soma:16km:error}. As the second order method, ETD2wave has the best performance among these three methods, which is with the smallest errors comparing to RK4. Since B-ETD2wave method is a model reduction method, it considers the reduced model and bears some extra reduction error.  Exponential Rosenbrock-Euler has larger errors than ETD2wave method, but smaller errors than B-ETD2wave.  
}

\begin{table}[ht!]
	\centering
	\begin{tabular}{rrrr}
		\hline
		&Rosenbrock-Euler& ETD2wave& B-ETD2wave\\ 
		\hline
		$h_1$ & 1.3128e-06 & 1.0066e-08  & 3.2326e-06 \\
		$\bu_1$ & 4.5907e-05 & 2.7242e-07  & 9.4878e-05 \\
		\hline
	\end{tabular}
	\caption{The relative $l_{\infty}$ errors of simulated layer thickness  and velocity  of the first layer on Day 15 by the parallel ETD methods.}\label{soma:16km:error}
\end{table}

To measure the parallel efficiency, we define $E_p=\frac{r\cdot T_r}{p\cdot T_p}$, where $p$ is the number of used cores  and $T_p$ is the associated CPU time, $r$ is the minimum number of cores considered in the test case and the corresponding CPU time $T_r$ is regarded as the reference. The simulation performances are reported in Tables \ref{soma:16km}-\ref{soma:4km}.

\begin{table}[ht!]
	\centering
\begin{tabular}{rrrrrrr}
	\hline
	\multirow{2}{*}{Cores}&\multicolumn{2}{c}{Rosenbrock-Euler}& \multicolumn{2}{c}{ETD2wave}& \multicolumn{2}{c}{B-ETD2wave}\\ \cmidrule(lr){2-3}\cmidrule(lr){4-5}\cmidrule(lr){6-7}
	&Time (s)&Efficiency &Time (s)&Efficiency &Time (s)&Efficiency\\ \hline
	8 & 194.64& - & 76.64& - &27.99 & - \\
	16& 100.14& 92\%& 43.11& 89\%&17.87 & 78\%   \\
	32& 84.82 & 58\%& 27.2& 70\% & 12.56& 56\%   \\
	64& 35.71 & 68\%&16.10 & 60\%&8.17 & 43\%   \\
   128& 18.57 & 66\%&10.57 & 45\%&6.61 & 26\%   \\
	\hline
\end{tabular}
\caption{CPU times and parallel efficiency  for the SOMA test case: 16 km resolution mesh.}\label{soma:16km}
\end{table}

\begin{table}[ht!]
	\centering
\begin{tabular}{rrrrrrr}
	\hline
	\multirow{2}{*}{Cores}&\multicolumn{2}{c}{Rosenbrock-Euler}& \multicolumn{2}{c}{ETD2wave}& \multicolumn{2}{c}{B-ETD2wave}\\ \cmidrule(lr){2-3}\cmidrule(lr){4-5}\cmidrule(lr){6-7}
	&Time (s)&Efficiency &Time (s)&Efficiency &Time (s)&Efficiency\\ \hline
	8&855.73& -   & 470.61& -   & 140.25& -   \\
	16&482.32& 89\% & 281.73& 84\%& 78.26& 90\%   \\
	32&343.71& 62\%& 216.39& 54\%& 51.67& 68\%\\
	64&173.84& 62\%& 93.15& 63\%& 26.47& 66\%\\
	128&92.20 & 58\%& 33.67& 87\%& 14.51 & 60\%\\
	256&51.95 & 51\%& 19.39& 76\%& 11.76 & 37\%\\
	\hline
\end{tabular}
\caption{CPU times and parallel efficiency for the SOMA test case: 8 km resolution mesh.}\label{soma:8km}
\end{table}

\begin{table}[ht!]
	\centering
\begin{tabular}{rrrrrrr}
	\hline
	\multirow{2}{*}{Cores}&\multicolumn{2}{c}{Rosenbrock-Euler}& \multicolumn{2}{c}{ETD2wave}& \multicolumn{2}{c}{B-ETD2wave}\\ \cmidrule(lr){2-3}\cmidrule(lr){4-5}\cmidrule(lr){6-7}
	&Time (s)&Efficiency &Time (s)&Efficiency &Time (s)&Efficiency\\ \hline
	16& 2316.33 & - & 1414.57 & - & 528.82 & -\\
	32& 1657.91 & 70\% & 1069.01 & 66\% & 364.62 & 73\%\\
	64& 737.64 & 79\% & 478.49 & 74\% & 157.13 & 84\% \\
	128& 369.09 & 78\% & 221.02 & 80\% & 53.99 & 122\%\\ 
	256& 188.15 & 77\% & 101.67 & 87\% & 29.01 & 114\%\\
	\hline
\end{tabular}
\caption{CPU times and parallel efficiency for the SOMA test case: 4 km resolution mesh.}\label{soma:4km}
\end{table}

It is seen that the parallel efficiency of the three methods increases as the resolution increases. A higher resolution mesh contains more degrees of freedom (vertices, edges, and cells), which results in more computing tasks to each core and  leads to the improvement on parallel efficiency. Because of some model reduction, B-ETD2wave achieves a better overall performance than the other two. For instance, when the minimum cores are used, the B-ETD2wave scheme is about five times faster than the exponential Rosenbrock-Euler and three times faster than ETD2wave. Since the exponential Rosenbrock-Euler requires more local computing on the Arnoldi process, its parallel efficiency is better than the other two methods. On the 4 km resolution mesh, its parallel performance  becomes worse than the other two methods. A potential reason is that we take the computing time of 16 processes as the reference. Exponential Rosenbrock-Euler has a better balance between the inter-process communication and local computing with 16 processes on the Cori. Later the inter-process communication wins over local computing, which can be seen from the overall decreasing efficiency. While for the other two methods, 16 processes are not their optimal task decomposition and thus  their performance keeps increasing and even gives super-linear speedups.

\subsection{The baroclinic eddies test case for the primitive equations with twenty layers}

\begin{figure}[!ht]
	\centering{
		\includegraphics[height=2.3in]{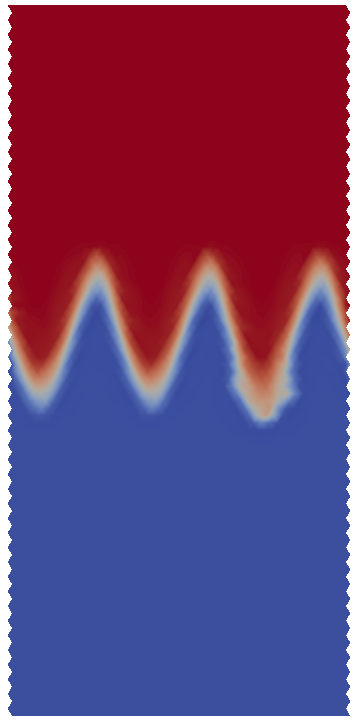}
		\includegraphics[height=2.3in]{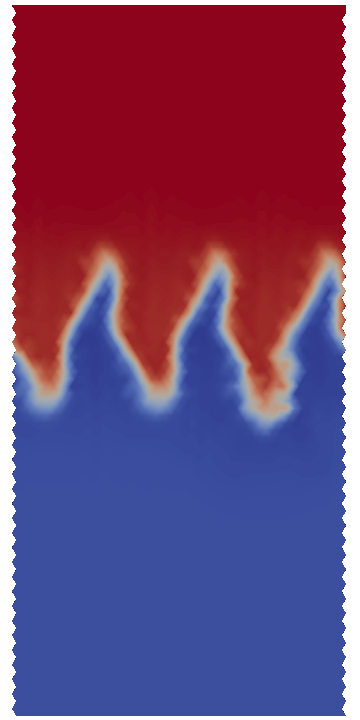}
		\includegraphics[height=2.3in]{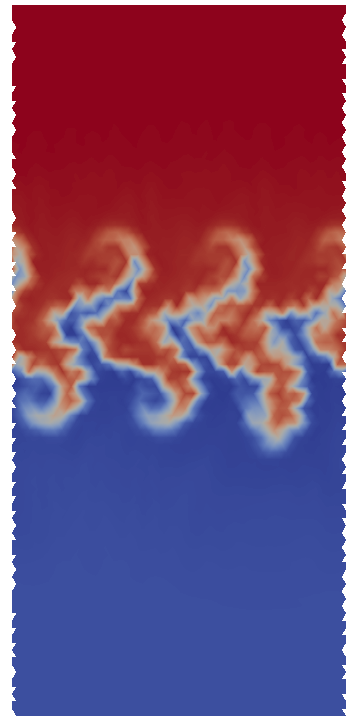}
		\includegraphics[height=2.3in]{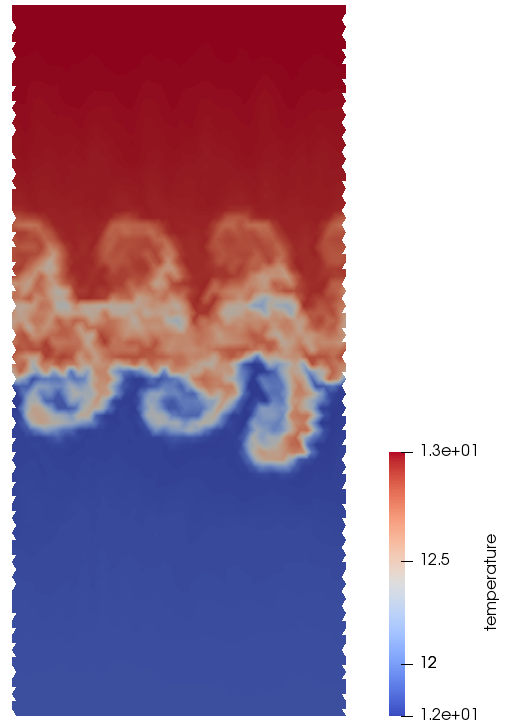}
	} 
\caption{Simulated temperatures by our ETD-SP2 with $\Delta t=60$ s for the baroclinic eddies test case. From left to right are the results for day 1, 5, 10 and 15.}\label{temp:etd}
\end{figure}

\begin{figure}[!ht]
\centering{
	\includegraphics[height=2.3in]{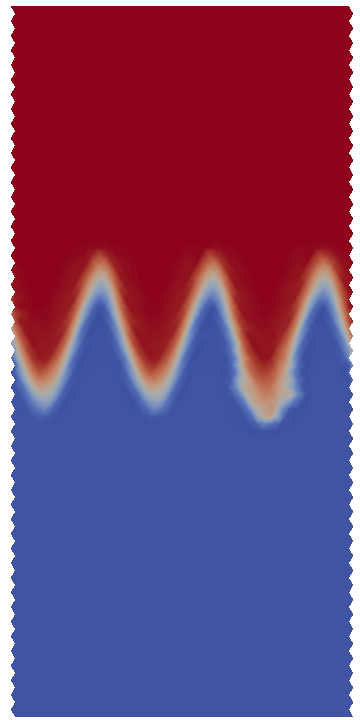}
	\includegraphics[height=2.3in]{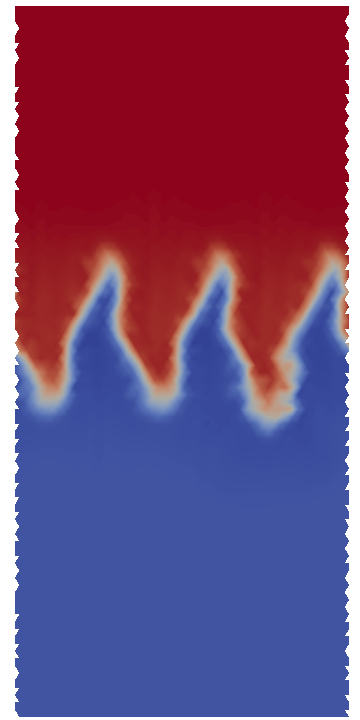}
	\includegraphics[height=2.3in]{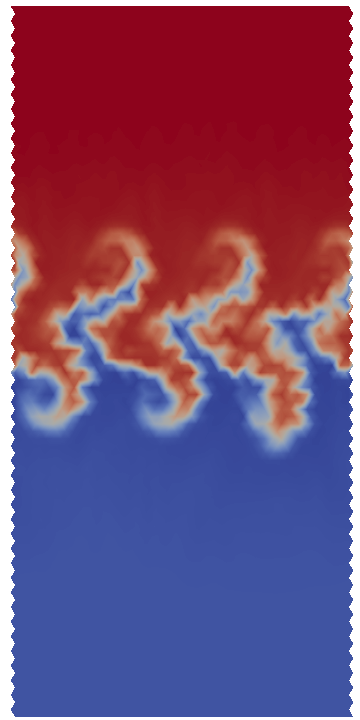}
	\includegraphics[height=2.3in]{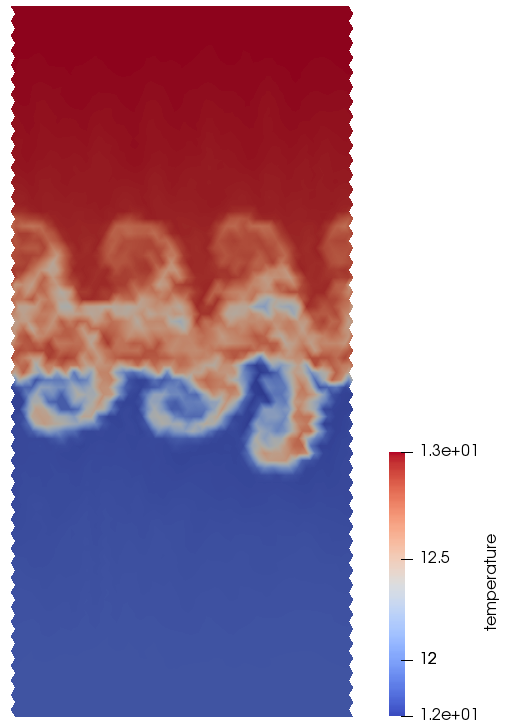}
}
\caption{Simulated temperatures by  SP2 in MPAS with $\Delta t=60$ s and $\Delta t_{\text{btr}}=4s$  for the baroclinic eddies test case. From left to right are the results for day 1, 5, 10 and 15.}\label{temp:sp2}
\end{figure}

\begin{figure}[!ht]
\centering{\hspace{0.6cm}
	\includegraphics[height=2.3in]{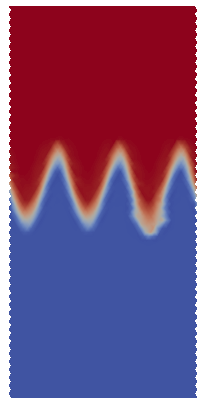}\hspace{-0.1cm}
	\includegraphics[height=2.3in]{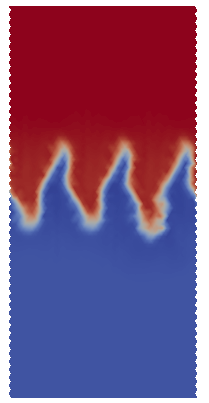}\hspace{-0.1cm}
	\includegraphics[height=2.3in]{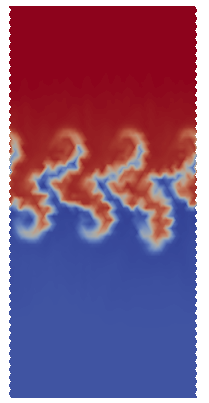}\hspace{-0.1cm}
	\includegraphics[height=2.3in]{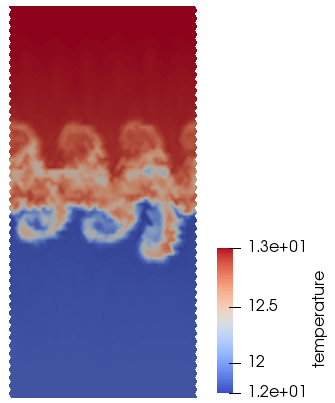}
} 
\caption{Simulated temperatures  by RK4 in MPAS with $\Delta t=15$ s  for the baroclinic eddies test case. From left to right are the results for day 1, 5, 10 and 15.}\label{temp:rk4}
\end{figure}

\begin{figure}[!ht]
	\centering{\hspace{0.6cm}
		\includegraphics[height=2.1in]{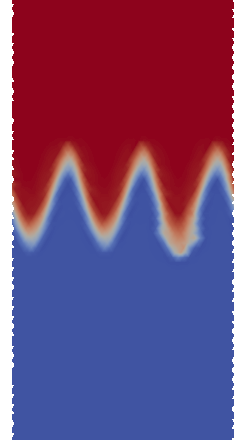}\hspace{-0.1cm}
		\includegraphics[height=2.1in]{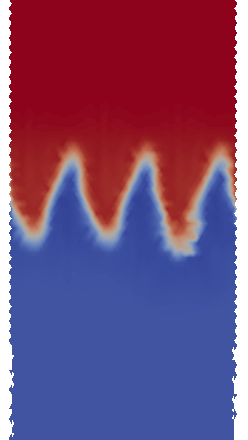}\hspace{-0.1cm}
		\includegraphics[height=2.1in]{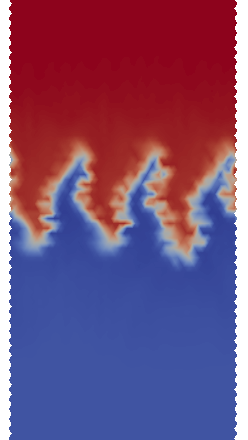}\hspace{-0.1cm}
		\includegraphics[height=2.1in]{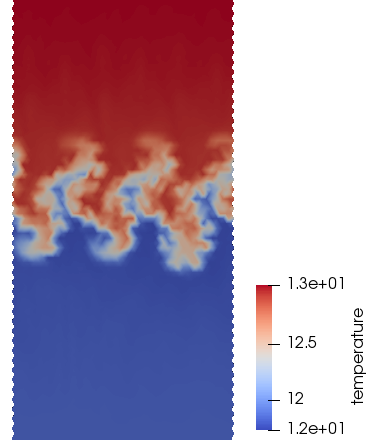}
	} 
	\caption{Simulated temperatures  by  ETD-SP1 with $\Delta t=60$ s  for the baroclinic eddies test case. From left to right are the results for day 1, 5, 10 and 15.}\label{temp:etd1}
\end{figure}

Next we consider a test case of the primitive equations with twenty layers from MPAS-Ocean \cite{ringler2013multi,petersen2018mpas} imported from \cite{ilicak2012spurious}. The domain consists of
a horizontally periodic channel of latitudinal extent 500 km and
longitudinal extent 160 km, with a flat bottom of 1000 m vertical depth. The channel is on a f-plane \cite{cushman2011introduction} with the Coriolis parameter $f=1.2\times10^{-4}$ s$^{-1}$. The initial temperature
 decreases downward in the meridional direction. A cosine shape temperature
perturbation with a wavelength of 120 km in the zonal direction is used to instigate the baroclinic
instability. We will use a 10-km-resolution SCVT mesh containing 3,920 cells, 11,840 edges, and 7,920 vertices, and  carry out 15 days simulation.
Considering it is a small scale computing, we run all numerical  experiments in  this test with 6 cores.
The maximum dimension of Krylov subspaces used in ETD methods is  set  to  be 25. 
The numerical results by our ETD-SP2,  SP2, and RK4 are shown in Figures \ref{temp:etd}, \ref{temp:sp2}, and \ref{temp:rk4} respectively, and all of them
perform very similarly. In addition, we test ETD-SP1 for comparison and present its results in Figure \ref{temp:etd1}. It is observed that the temperature obtain by
ETD-SP1 is less diffusive than that of ETD-SP2 and RK4. The inaccuracy is mainly caused by the sub-sampling influence discussed in Section \ref{subsect:etdsp}.


To test these methods' accuracy, we choose the classical RK4 with $\Delta t=1$ s as the benchmark solution and compare the errors of the surface temperature at $1$ hour. 
The relative $l_{\infty}$ errors of simulated surface temperature by ETD-SP2 and SP2 are listed in Table \ref{barochannel:errortest}. 
Since we develop our code in MPAS-Ocean by replacing its barotropic sub-stepping part with the ETD method, we compare the time cost of computing the barotropic velocity (Stage 2 in Section \ref{subsect:etdsp}) with four cores. The results are listed in Table \ref{barochannel:errortest}, in which $\Delta t$ is the time step size for both baroclinic and barotropic modes in our ETD-SP2 method, $\Delta t$ and $\Delta_{\text{btr}} t$ are the ones for the baroclinic and  barotropic modes, respectively, in SP2. From Table  \ref{barochannel:errortest}, we observe that both ETD-SP2 and SP2 only achieves the first-order accuracy in time. It is probably because one advances the barotropic mode over $2\Delta t$ to avoid sub-sampling its high frequency in the current MPAS-Ocean numerical model. Since MPAS is based on the MPDATA method, it suggests using the velocity at the middle time level, $t_{n+1/2}$, to solve the transport equation. But we input the barotropic velocity at $t_{n+1}$ instead of $t_{n+1/2}$, so it might degrade the accuracy. On the other hand, we can see that our ETD-SP2 spends less time than SP2 (about a half of the time cost of SP2) thus is more efficient in this case. This gain comes from the single step at solving the barotropic velocity, while the original sub-stepping method in SP2 needs multiple small time steps. 


\begin{table}[ht!]
	\centering
	\begin{tabular}{rrrrrrr}
		\hline
		\multirow{2}{*}{$\Delta t(\Delta_{\text{btr}} t)$}&\multicolumn{3}{c}{ETD-SP2}& \multicolumn{3}{c}{SP2}\\ 
		\cmidrule(lr){2-4}\cmidrule(lr){5-7}
		& Error&Rate&Time (s)&Error&Rate&Time (s)\\ \hline
		60s(8s)&6.1029e-05&-&0.37&6.1164e-05&-&0.69\\
		30s(4s)&3.0709e-05&0.99&0.75&3.0776e-05&0.99&1.37\\
		15s(2s)&1.4884e-05&1.04&1.46&1.4927e-05&1.04&2.75\\
		8s(1s) &7.4326e-06&1.00&2.86&7.4590e-06&1.00&5.40\\
		\hline
	\end{tabular}
	\caption{The relative $l_{\infty}$ errors and convergence rates of simulated surface temperature and the corresponding CPU times  by ETD-SP2 and SP2 with 6 cores for the baroclinic eddies test case.}\label{barochannel:errortest}
\end{table}


\subsection{The global ocean test case for the primitive equations with sixty layers}
\begin{figure}[!ht]
	\centerline{
		\includegraphics[height=2.8in]{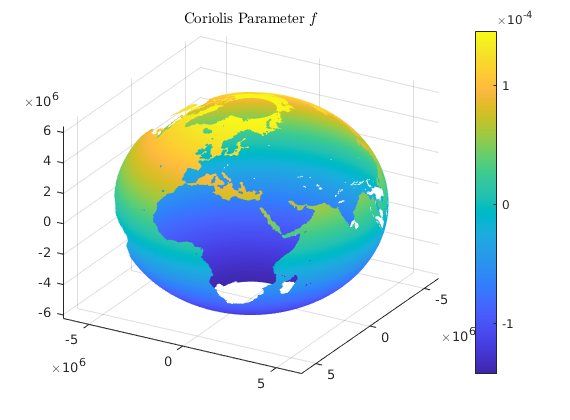}
	} 
\caption{The distribution of Coriolis parameter value.}\label{EC60to30:f}
\end{figure}

\begin{figure}[!ht]
	\centerline{
		\includegraphics[height=2.3in]{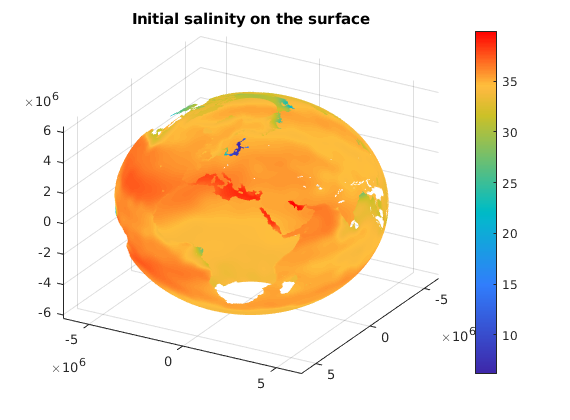}
		\includegraphics[height=2.3in]{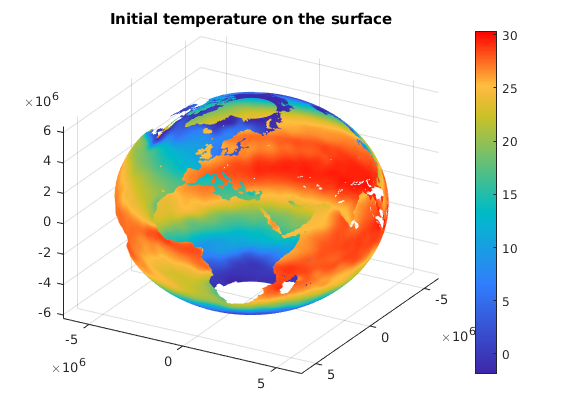}	
	} 
\caption{Initial values of the salinity and temperature.}\label{EC60to30:init}
\end{figure}

\begin{figure}[!ht]
	\centerline{
		\includegraphics[height=2.3in]{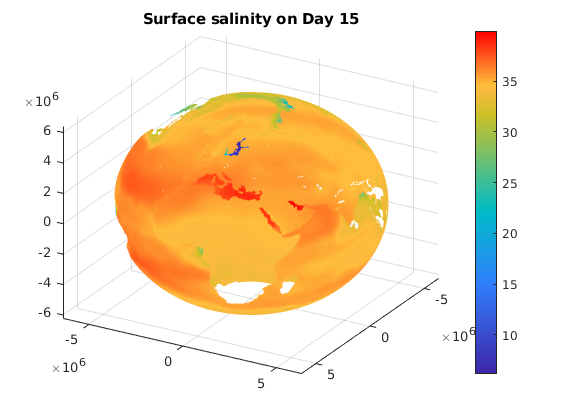}
		\includegraphics[height=2.3in]{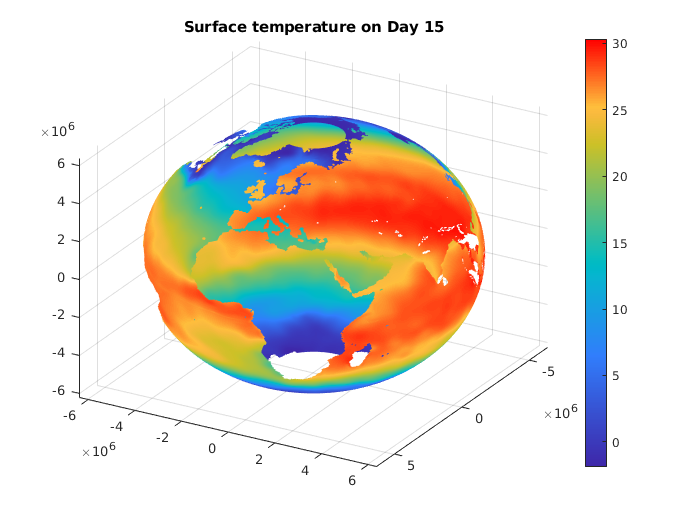}	
	} 
	\caption{Simulated surface salinity and temperature on Day 15 by ETD-SP2 with $\Delta t=60$ s for the global ocean test case.}\label{EC60to30:d15}
\end{figure}

\begin{figure}[!ht]
	\centerline{
		\includegraphics[height=2.3in]{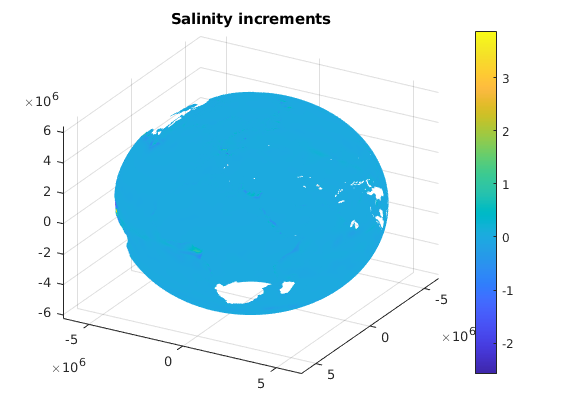}
		\includegraphics[height=2.3in]{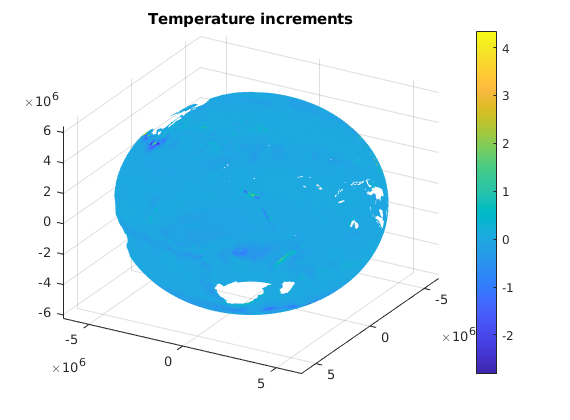}	
	} 
	\centerline{
		\includegraphics[height=2.3in]{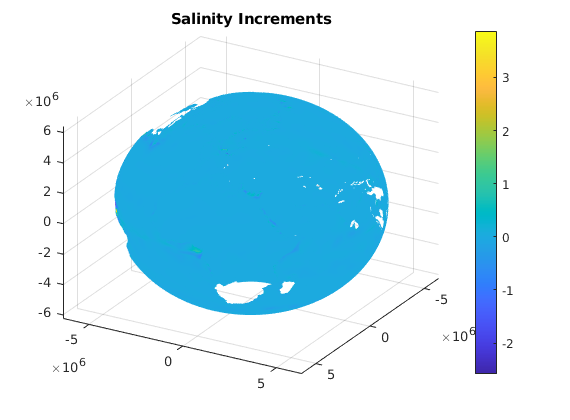}
		\includegraphics[height=2.3in]{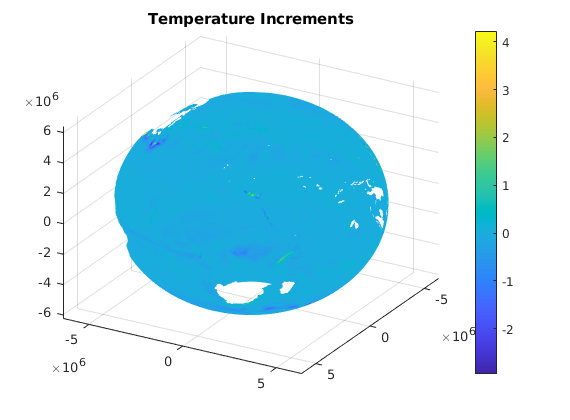}	
	} 
	\caption{Simulated increments of the surface salinity  and temperature  on Day 15 by ETD-SP2 (top panel) and RK4 (bottom panel) with $\Delta t=60$ s.}\label{EC60to30:d15:diff:etd-rk4}
\end{figure}

In this part, let us consider a test with global real-world configuration. The mesh provided by MPAS-Ocean, EC60to30, varies from 30 km resolution at the equator and poles to 60 km resolution at the mid-latitudes and uses 60 vertical layers. 
The grid contains 235,160 cells, 714,274 edges, and 478,835 vertices on each layer. The distribution of  the Coriolis parameter $f$ is shown in Figure \ref{EC60to30:f}. The initial conditions for temperature and salinity are interpolated from the Polar Science Center Hydrographic Climatology, version 3 \cite{steele2001phc} that are presented in Figure \ref{EC60to30:init}. We run the 15 days simulation using  ETD-SP2 with $\Delta t=60$ s and show the resulting surface temperature and salinity in Figure \ref{EC60to30:d15}. The increments of the two states generated by ETD-SP2 and RK4 with $\Delta t=60$ s are shown and compared in Figures \ref{EC60to30:d15:diff:etd-rk4}, and the results show that they perform very similarly.

In addition, we test the parallel efficiency of ETD-SP2 and SP2 by running 1 hour simulation and evaluating $E_p$ as done in Section \ref{soma:test}. To obtain the comparable accuracy, we again set the maximum dimension of Krylov subspace to be 25 in  ETD-SP2 and take the time step $\Delta t=60$ s. In SP2, we take the same $\Delta t =60$ s for the baroclinic mode and $\Delta t_{btr}=1$ s for the  baratropic mode. Comparing with the benchmark solution of RK4 with $\Delta t=5$ s, the  $l_\infty$ error of simulated surface temperature by  ETD-SP2 and SP2 are 7.5515e-05 and 7.5773e-05, respectively. Their computation  times with different number of cores  up to 256 are reported in Table \ref{etd:parallefficiency}, from which we observe that ETD-SP2 costs less computational times than SP2 in these cases (i.e., when the number of used cores is not very large), but it has relatively  lower parallel efficiency. That is partially due to the fact that ETD-SP2 needs more inter-process communication due to the Arnoldi process, its orthogonalization requires several matrix-vector multiplications and evaluate the norm of the resulting vectors. Therefore, its efficiency decreases as the number of cores increases.  SP2 only needs the inter-process communication to update the values on the halo cells at the end of each sub-time-stepping, which gives  super-linear speedups again.

\begin{table}[ht!]
	\centering
	\begin{tabular}{crrrr}
		\hline
		\multirow{2}{*}{Cores}&\multicolumn{2}{c}{ETD-SP2}&  \multicolumn{2}{c}{SP2}\\ 
		\cmidrule(lr){2-3}\cmidrule(lr){4-5}
		& Time (s)&Efficiency&Time (s)&Efficiency\\ \hline
		16& 44.44 & - & 225.61 & - \\
		32& 27.72 & 80.2\% & 126.01 & 89.5\% \\
		64& 16.68 & 66.6\% & 65.61 & 86.0\% \\
		128& 11.33 & 49.0\% & 27.90 & 100.9\% \\
		256& 6.85 & 40.6\% &11.59 & 121.4\% \\
		\hline
	\end{tabular}
	\caption{CPU times and parallel efficiency of ETD-SP2 and SP2 for the global ocean test case.}\label{etd:parallefficiency}
\end{table}

\section{Conclusions}\label{sect:conclusion}
In this paper, we have fulfilled and test the parallel implementation of several exponential time differencing methods for simulating the ocean dynamics,
governed by either the multi-layer SWEs or the multi-layer primitive equations.
Since  B-ETD2wave is developed on a reduced layered model, it costs less computations than  exponential Rosenbrock-Euler and ETD2Wave in solving multi-layer SWEs. We also designed a new ETD-SP2 method for solving the multilayer primitive equations. After splitting the oceanic motion into the baroclinic and barotropic components, we use the forward Euler scheme for the baroclinic mode and the exponential Rosenbrock-Euler scheme for the barotropic mode in the current 
MPAS-Ocean framework. Several standard numerical tests are performed and the comparison  results demonstrate a great potential of applying the parallel ETD methods in simulating real-world geophysical flows.

Since the tracers' equations, like temperature and salinity, are the advection equations, MPAS-Ocean adopts the method proposed in \cite{higdon2005two}, which is to utilize the MPDATA method and transport the tracers by the advective velocity at the intermediate time $t_{n+1/2}$. Smolarkiewicz and Margolin \cite{smolarkiewicz1998mpdata} designed the MPDATA method based on a general advection equation without considering the sub-sampling issue mentioned in \cite{mit2004}. As a future work, we will carefully design the new MPDATA method with the splitting transport velocities at different time levels to improve the temporal accuracy.

\section*{Acknowledgement}
 This work was supported by the U.S. Department of Energy, Office of Science, Office of Biological and Environmental Research through Earth and Environmental System Modeling and Scientific Discovery through Advanced Computing programs under grants DE-SC0020270 and DE-SC0020418, and the National Science Foundation under grant DMS-1818438. This research used the resources of National Energy Research Scientific Computing Center (NERSC), a U.S. Department of Energy Office of Science User Facility operated under Contract No. DE-AC02-05CH11231.

\bibliographystyle{abbrv}
\bibliography{ParallelETDOcean}
\end{document}